\documentclass[aps,pra,superscriptaddress,twocolumn,showpacs,noeprint]{revtex4-2}
\usepackage{graphicx,amsfonts,times,bm,amsmath,verbatim,color,array}

\usepackage{natbib}
\usepackage{graphicx,amssymb,amsmath,ifthen,braket,xcolor}
\usepackage{bm}

\begin{document}

\title{Quantifying Entanglement in Cluster States Built with Error-Prone Interactions}

\author{Zhangjie Qin}
\affiliation{Department of Physics, Virginia Tech, Blacksburg, Virginia 24061, USA}

\author{Woo-Ram Lee}
\affiliation{Department of Physics, Virginia Tech, Blacksburg, Virginia 24061, USA}

\author{Brian DeMarco}
\affiliation{Department of Physics and IQUIST, University of Illinois at Urbana-Champaign, Urbana, IL 61801, USA}

\author{Bryce Gadway}
\affiliation{Department of Physics and IQUIST, University of Illinois at Urbana-Champaign, Urbana, IL 61801, USA}

\author{Svetlana Kotochigova}
\affiliation{Department of Physics, Temple University, Philadelphia, PA 19122, USA}

\author{V.W. Scarola}
\email[Email address:]{scarola@vt.edu}
\affiliation{Department of Physics, Virginia Tech, Blacksburg, Virginia 24061, USA}

\begin{abstract}
\noindent

Measurement-based quantum computing is an alternative paradigm to the circuit-based model. This approach can be advantageous in certain scenarios, such as when read-out is fast and accurate, but two-qubit gates realized via inter-particle interactions are slow and can be parallelized to efficiently create a cluster state. However, understanding how two-qubit errors impact algorithm accuracy and developing experimentally viable approaches to characterize cluster-state fidelity are outstanding challenges. Here, we consider one-dimensional cluster states built from controlled phase, Ising, and XY interactions with slow two-qubit error in the interaction strength, consistent with error models of interactions found in a variety of qubit architectures. We detail an experimentally viable teleportation fidelity that offers a measure of the impact of these errors on the cluster state. Our fidelity calculations show that the error has a distinctly different impact depending on the underlying interaction used for the two-qubit entangling gate. In particular, the Ising and XY interactions can allow perfect teleportation through the cluster state even with large errors, but the controlled phase interaction does not.  Nonetheless, we find that teleportation through cluster state chains of size $N$ has a maximum two-qubit error for teleportation along a quantum channel that decreases as $N^{-1/2}$. To enable construction of larger cluster states, we design lowest-order refocusing pulses for correcting these slow errors in the interaction strength. Our work generalizes to higher-dimensional cluster states and sets the stage for experiments to monitor the growth of entanglement in cluster states built from error-prone interactions.
\end{abstract}


\maketitle

\section{Introduction}

Cluster states are entangled quantum many-body states of matter that can serve as resource states for  measurement-based quantum computing (MBQC) \cite{Raussendorf2001b,Raussendorf2003}.  The MBQC formalism shows that any quantum algorithm can be executed with properly arranged measurements of a cluster state.  Cluster states are also symmetry protected topological phases  \cite{Doherty2009,Miyake2010,Else2012c,Azses2020}.  
As such, certain Hilbert space sectors of cluster states offer robust routes to store and process information.  

MBQC with cluster states can offer advantages over the usual circuit-based approach to quantum information processing.  In systems with slow and possibly error-prone two qubit gates, the MBQC formalism front-loads the burden of executing two-qubit gates to the initial phase of the algorithm.  Furthermore, all two-qubit gates can be run in parallel (at the same time) to quickly build the cluster state. Single-qubit measurements then execute the algorithm and therefore avoid the later use of two-qubit gates.  The MBQC approach works best in systems with a large number of available qubits that can be entangled in parallel into a cluster state while allowing subsequent fast and accurate single-qubit measurements.   

Observable properties of cluster states depend on the level of entanglement left over after error-prone gates are used to create them. Characterization of entanglement in an $N$-qubit many-body state with full quantum state tomography has been done with small cluster states built from entangled photons \cite{Walther2005,Schwartz2016} and with ions \cite{Lanyon2013}.  But, in general, full quantum state tomography
scales exponentially with $N$ and is well known to be prohibitive.

Teleportation fidelity offers a route to measure entanglement between qubits \cite{VanEnk2007}.  Teleportation measures of fidelity typically envision operations on a pair of well-separated qubits.  But teleportation \emph{through} a cluster state is different.  The process was originally introduced  \cite{Raussendorf2001b,Raussendorf2003} as  a logical identity gate where information encoded in one end of the cluster state is teleported via measurement to the other side.  It passes information within a many-body state of matter in its entirety, engaging all qubits in the measurement.  The process of measurement-induced teleportation relies on entanglement in the cluster state itself and is therefore an implicit test of cluster state entanglement.  As a result, several theoretical studies have examined the interplay of teleportation and errors in cluster states \cite{Tame2005,TAME2006,Alexander2016} where thresholds \cite{Horodecki1999,Braunstein2001,Bose2003,Paternostro2005} for teleportation along a quantum channel (as opposed to classical transmission of information) were studied.  Recent works have used teleportation to test the ability of symmetry-protected topological order in cluster states to protect against errors (see, e.g., Refs.~\cite{Else2012c, Azses2020}).

Cluster states are built from two-qubit entangling gates.  The most compact gate set used to construct cluster states utilizes the controlled phase gate  \cite{Raussendorf2001b,Raussendorf2003}.  This two-qubit gate has been constructed as a composite gate (built from other gates or operations) between ion \cite{Cirac1995,Bruzewicz2019}, superconducting \cite{Strauch2003,Dicarlo2009}, and photonic \cite{Walther2005} qubits, which show considerable promise in MBQC \cite{Nielsen2004,Kok2007}.  But the controlled phase gate also arises from physical interactions between particles used as qubits, e.g., neutral atoms with Rydberg excitations \cite{Jaksch2000,Lukin2001,Saffman2010,Saffman2005,Morgado2021} or controlled collisions in optical lattices  \cite{Jaksch1999,Mandel2003}, that do not rely on a composite two-qubit gate construction. 
Other two-qubit entangling interactions, the Ising interaction \cite{Raussendorf2001b,Raussendorf2003} and the XY (conditional phase flip) interaction \cite{Tanamoto2009a} can also be used to efficiently create cluster states directly from inter-particle interactions.  The Ising interaction  characterizes the M{\o}lmer-S{\o}renson gate between ion qubits \cite{Molmer1999,Leibfried2003,Ballance2016,Gaebler2016,Bruzewicz2019}, the interactions between NMR qubits \cite{Vandersypen2005}, and the interaction between superconducting charge qubits \cite{Krantz2019,You2005}.  And the XY interaction characterizes 
the entangling interaction between qubits formed from rotational states of polar molecules \cite{DeMille2002,Yan2013a,Ni2018}, quantum dots in cavities \cite{Imamo1999}, and other types of superconducting qubits \cite{You2005}. 

We consider the intertwined obstacles of growing and diagnosing entanglement in cluster states from the point of view of challenges and goals in the laboratory setting.  Moving from a few qubits to several requires a growth and measurement protocol that avoids costly $N$-qubit quantum state tomography along with a roadmap to mitigate errors.  Adding to the complexity, we find that the impact of errors on entanglement depends on the physical two-qubit gate used to build the cluster state.  To tackle these obstacles, we study cluster states built from three types of error-prone interactions: controlled phase, Ising, and XY, because these interactions are commonly used to characterize the physical (native) interaction between particles in many architectures and they lead directly to well known entangling gates.  

While all of the above architectures have a variety of error and noise sources, here we focus on slow errors in the native-two qubit coupling.  Two qubit gates tend to have a lower fidelity than the single-qubit gates \cite{DeLeon2021}.  We therefore ignore single qubit errors as a first approximation.  Furthermore, in many qubit platforms, particularly those formed from atoms and molecules, fluctuations in control fields, e.g., laser power or orientation, perturb  interactions thereby causing slow two-qubit gate errors.  For example, consider two qubits defined by rotational states of polar molecules trapped in optical tweezers.  Slow relative perturbations to the position of the laser focal point will perturb the dipolar interaction and therefore lead to two-qubit gate errors \cite{DeMille2002,Ni2018}. In general, we consider two-qubit interactions of strength $J$, applied for a time $t$, with unknown errors $\varepsilon$, such that the gate strength is perturbed by an error: $Jt\rightarrow Jt (1 +\varepsilon)$, where $\varepsilon$ does not depend on time for a single application of a two-qubit gate (errors are slow on time scales of the interaction energy).  We allow for the possibility that $\varepsilon$ changes from gate to gate. 

We construct protocols for experiments to use teleportation fidelity as a low-cost, $\mathcal{O}(N)$, operation to benchmark the impact of errors on entanglement in cluster states \cite{Tame2005,TAME2006}.  We numerically test teleportation fidelity benchmarking on cluster states to predict what experiments should see.  We focus on cluster state chains but our results apply to two and three dimensions without loss of generality because analogues of the logical identity gate apply to higher dimensions (e.g., the teleportation protocol discussed here leads to a logical SWAP operation when applied to a two-dimensional cluster state \cite{Raussendorf2003}). We study the impact of errors in constructing cluster states with the controlled phase interaction \cite{Tame2005,TAME2006}, the Ising interaction, and the XY interaction.  The errors shorten the cluster state chain lengths that allow teleportation along a quantum channel.  To allow for longer chain lengths, we also construct routes to refocus two-qubit gate errors.

We find that, even though the protocols for constructing the cluster states with these different interactions are very similar, the impact of errors are quite different.  The cluster state built with the controlled phase interaction requires the fewest number of gates but we will see that errors here are hardest to correct with common refocusing schemes \cite{Jones2003,Hill2007,Tomita2010a}.  The Ising and XY interaction require additional gate overhead to construct the cluster state but we find two key differences in comparison to the controlled phase interaction. First, we find that the errors in Ising and XY interactions still allow perfect teleportation of certain qubit states along the chain because of symmetry in the cluster state.  Second, we find that refocusing the Ising and XY interactions requires fewer gates.  Overall, the protocols we construct predict that incrementally growing and benchmarking entanglement with teleportation fidelity should be doable but we also find trade-offs in constructing cluster states with controlled phase interactions as opposed to the Ising or XY interaction.

The paper is organized as follows.  In Sec.~\ref{sec_teleport}, we revisit the MBQC identity gate \cite{Raussendorf2001b,Raussendorf2003} as a teleportation fidelity \cite{Tame2005,TAME2006}.  Sec.~\ref{sec:controlled} then discusses how to construct the cluster state with the controlled phase interaction.  We show numerical results for what an experiment should be able to observe in teleportation fidelity with error in the controlled phase interaction energy.  Throughout the paper, we assume that errors slow on the time scale of a single two-qubit gate are the only source of error.  Secs.~\ref{sec:Ising} and \ref{sec:XY} do the same but for the Ising and XY interactions, respectively.   Sec.~\ref{sec:refocusing} constructs minimal refocusing schemes to mitigate the impact of the two-qubit interaction error.  We summarize in Sec.~\ref{sec:summary}.

\section{Cluster State Teleportation, Fidelity, and Entanglement}
\label{sec_teleport}

\begin{figure}
\begin{center}
\includegraphics[width=0.46\textwidth]{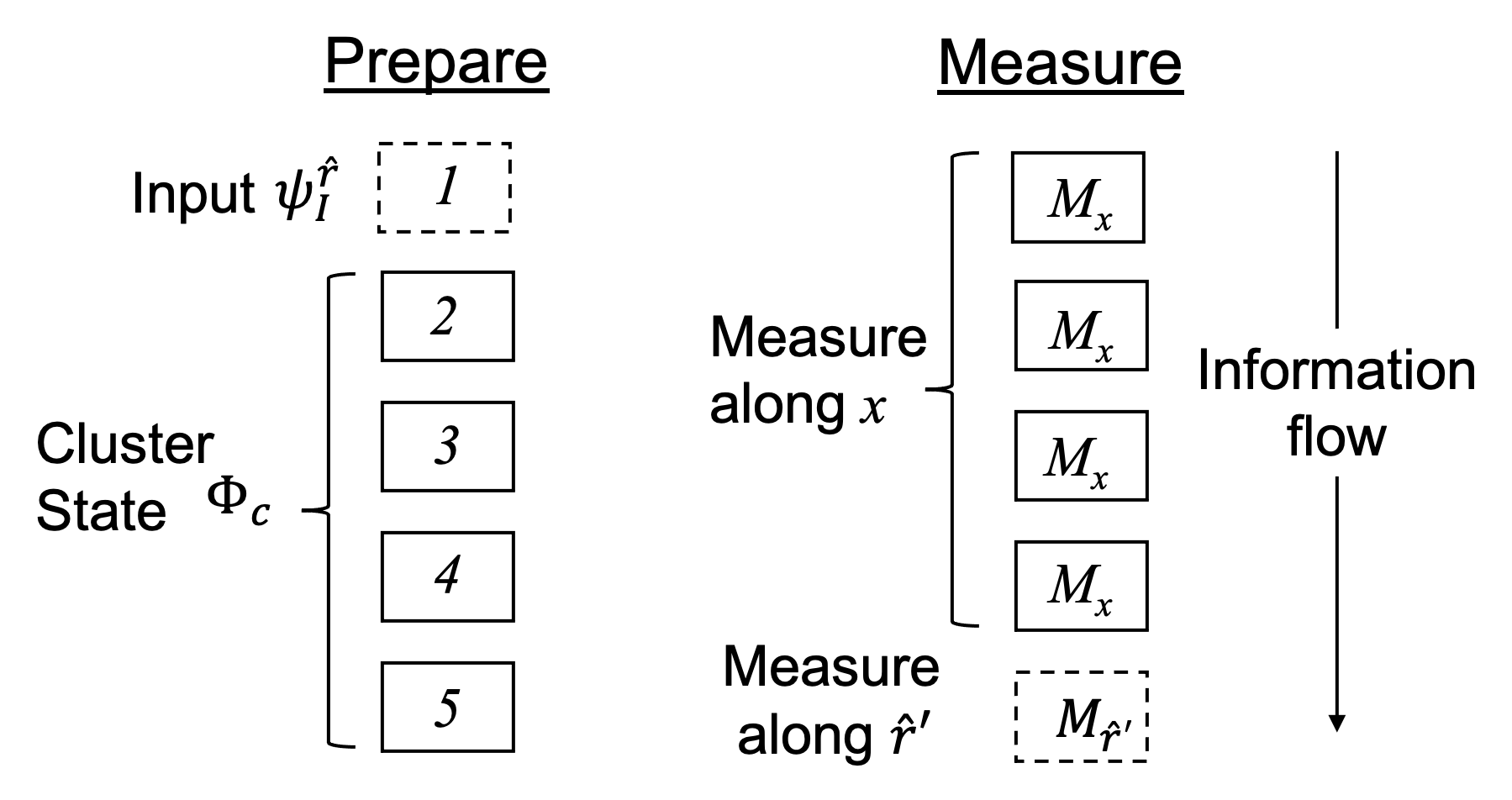}
\end{center}
\caption{Schematic of two stages of the fidelity measurement shown for 5 qubits.  The first column depicts preparation of qubit 1 defining the input state,  $\vert \psi_{\text{I}}^{\hat{r}} \rangle$, where information is encoded in the qubit orientation, $\hat{r}$.  This qubit is entangled with the remaining qubits (2-5) that define a cluster state $\vert \Phi_C\rangle$.  The resulting state $\vert \psi_{\text{I}}^{\hat{r}}, \Phi_{\text{C}}  \rangle$ is then measured.  The second column depicts the measurement stage where measurement along the qubit-$x$ direction for qubits 1 through 4 effectively moves the information encoded in qubit 1 to qubit 5.  Quantum state tomography on qubit 5 allows reconstruction of an output state, $\vert \psi_{\text{O}} \rangle$, that is, in the absence of error, identical to the input, $\vert \psi_{\text{I}}^{\hat{r}} \rangle$, up to a matrix, $U_{\Sigma}$, defined from measurement results of qubits 1 to 4.  An error-free cluster state teleports information along a quantum channel from qubit 1 to qubit 5.  Sufficiently strong gate errors will destroy entanglement in the cluster state which can be detected in degraded teleportation.  The information recorded in the measurements is used offline to construct the fidelity $\mathcal{F}_{\hat{r}}$ of the cluster state, where $\mathcal{F}_{\hat{r}}> 2/3$ guarantees a quantum channel.
}
\label{fig_schematic_flow}
\end{figure}

We define the cluster state fidelity using the MBQC identity gate \cite{Raussendorf2001b,Raussendorf2003,Tame2005,TAME2006}. The MBQC identity gate relies on measurements to teleport information from an input qubit on one end of the cluster state chain to the other. Figure~\ref{fig_schematic_flow} shows a schematic of the process used to measure the fidelity. By using this process as a fidelity, we diagnose the entanglement in the cluster state, in so far as it can be used in teleportation. The fidelity defined here generalizes to higher dimensional cluster states, but we will focus on one dimension because it requires the lowest number of qubit resources.

To define the fidelity we denote the one-dimensional cluster state wavefunction containing $N-1$ qubits by $\vert \Phi_\text{C} \rangle $.  We then consider a single input qubit, $\vert \psi_{\text{I}}^{\hat{r}} \rangle$, that is prepared and entangled with $\vert \Phi_\text{C} \rangle $ to create a combined $N$-qubit state: $\vert \psi_{\text{I}}^{\hat{r}}, \Phi_{\text{C}}  \rangle$, where $\hat{r}$ is a unit vector pointing to a location on the Bloch sphere.  The fidelity measures the ability of measurements performed on the cluster state to move information stored in ${\hat{r}}$ along the chain.

To move the information along the chain, a series of measurements act on $\vert \psi_{\text{I}}^{\hat{r}}, \Phi_{\text{C}}  \rangle$ to effectively teleport.  The initial $N$-qubit wavefunction is measured to create a \emph{post}-measurement state.  Assuming measurement outcomes are recorded, we can write the final outcome of measurements as:
\begin{align}
\mathcal{P}(M_{\hat{r}'})\mathcal{P}(M_{\hat{x}}^{N-1}) \vert  \psi_{\text{I}}^{\hat{r}}, \Phi_{\text{C}}  \rangle,
\end{align}
where measurements along the qubit-$x$ direction on the first $N-1$ qubits are defined by the projectors:
$\mathcal{P}(M_{\hat{x}}^{N-1})=\prod_{j=1}^{N-1} (\sigma^0_{j}+(-1)^{s_j} \sigma^x_{j})/2$,
and the measurement along the qubit-$\hat{r}'$ direction on the final $N^{\text{th}}$ qubit is defined by the projector:
$\mathcal{P}(M_{\hat{r}'})=(\sigma^0_{N}+(-1)^{s_{N}} \hat{r}'_{N}\cdot\hat{\sigma}_{N})/2$.  Here, the integers $s_j =0,1$ are the eigenvalues that result from the measurement of the $j^{\text{th}}$ qubit and $\hat{\sigma}_{i}=(\sigma^x_i,\sigma^y_i,\sigma^z_i)$ is a vector of the usual Pauli matrices for qubit $i$.  In what follows, $\sigma^0_i$ denotes the identity matrix for qubit $i$.

Repeating this process allows quantum state tomography on the final $N^{\text{th}}$ qubit to yield the single-qubit output density matrix, $\rho_\text{O}$.  The output qubit orientation should be identical to the chosen input orientation $\hat{r}$, so that we retrieve $(\sigma^0+\hat{r}\cdot\hat{\sigma})/2$ for $\rho_\text{O}$.  But there are two caveats.  The first caveat is that errors lead to a reorientation of the output qubit.  The role of two-qubit errors will be discussed in later sections.

The second caveat is well known in usual teleportation schemes \cite{VanEnk2007}.  Even in the absence of errors, the measurement process inherently randomizes the measurement basis needed to interpret output on the last qubit.  A classical channel is needed to feedforward and interpret the output measurement.  MBQC uses measurement outcomes to construct a byproduct matrix (made from a combination of Pauli operations) that gives the correct basis in which to interpret the measurements on the final qubit.  Specifically, we record all measurement outcomes, $s_i$ on the first $N-1$ qubits.  The $s_i$ are then used to construct the appropriate byproduct matrix, $U_{\Sigma}$.  For odd $N$, we have \cite{Raussendorf2003}:
\begin{align}
U_{\Sigma}\equiv\prod_{i=1}^{(N-1)/2}(\sigma^x_{2i})^{s_{2i}}(\sigma^z_{2i-1})^{s_{2i-1}},
\label{eq_Usigma_CP}
\end{align}
where the product runs over all  measurements except the last qubit.  Here we see that the measurement outcomes, $s_i$, feedforward into interpretation of the quantum state tomography data on the $N^\text{th}$ qubit.  

Once the results of all measurements are recorded offline, we can construct the cluster state fidelity as the MBQC identity gate \cite{Raussendorf2001b,Raussendorf2003,Tame2005,TAME2006}.  As above, we assume that the input density matrix is a pure state, so that   $\rho_\text{I}^{\hat{r}}= \vert\psi_{\text{I}}^{\hat{r}}\rangle \langle \psi_\text{I}^{\hat{r}} \vert$ and define the fidelity to be:
\begin{align}
\mathcal{F}_{\hat{r}} \equiv 
\langle \psi_\text{I}^{\hat{r}} \vert U_{\Sigma}   \rho_{\text{O}} U_{\Sigma} \vert \psi_\text{I}^{\hat{r}} \rangle, 
\label{eq_fidelitysingle}
\end{align}
where the $U_{\Sigma}$ is defined in the limit of no error, Eq.~\eqref{eq_Usigma_CP}, to rotate our measurement outcome $\rho_{\text{O}}$ so that the output qubit orientation perfectly aligns with ${\hat{r}}$ if there is no error.  $\rho_{\text{O}}$ can, in general, describe a mixed state, but in what follows we will focus only on slow two-qubit gate errors.  In the presence of slow two-qubit error, the system stays closed. This leaves the input and output states as pure states, i.e., $\rho_\text{O}\rightarrow \vert \psi_{\text{O}} \rangle\langle \psi_{\text{O}} \vert $.  We then have $\mathcal{F}_{\hat{r}} \rightarrow  \vert \langle \psi_{\text{I}}^{\hat{r}} \vert U_{\Sigma}  \vert \psi_{\text{O}}  \rangle \vert ^2$.  

Equation~\eqref{eq_fidelitysingle} shows that if the output qubit is aligned along $\hat{r}$, the process used to define $\mathcal{F}_{\hat{r}}$ effectively executes the logical identity gate in MBQC and we have $\mathcal{F}_{\hat{r}}=1$ regardless of our choice of $\hat{r}$.  In such a case, the cluster state is perfectly entangled and teleportation occurs along a quantum channel. But in the presence of two-qubit error, the output qubit will not be aligned along $ {\hat{r}} $ and we might have  $\mathcal{F}_{\hat{r}}<1$ for some or all choices of $\hat{r}$.  $\mathcal{F}_{\hat{r}}$ therefore measures the deviation from the logical identity gate induced by error.    

The minimum fidelity aides in quantifying the extent to which error will degrade the fidelity. To benchmark over a random sample of input qubit orientations, we repeat the procedure for various input qubit directions $\hat{r}$.  Some initial orientations ${\hat{r}} $ have a lower fidelity than others.  The entanglement in the cluster state is ensured to contain a quantum channel \cite{Horodecki1999,Braunstein2001,Bose2003,Paternostro2005} if $\mathcal{F}_{\hat{r}}>2/3$. For $2/3 \geq \mathcal{F}_{\hat{r}}>1/2$, the channel could be either quantum or classical.  For  $\mathcal{F}_{\hat{r}} \leq 1/2$, entanglement in the cluster state has degraded to a point where there is only a classical channel.  The best measure of entanglement is then the minimum fidelity found for the worst case $\hat{r}$, defined to be $\text{Min}({\mathcal{F}})$.

The maximum fidelity, by contrast, helps locate protected channels.  It might be possible to find protected routes of teleportation that are, due to the interplay of symmetry in the cluster state \cite{Else2012c,Azses2020} and the error model, less sensitive to errors than other routes. In Sec. \ref{sec:Ising} and \ref{sec:XY}, we show that certain input orientations ${\hat{r}}$ allow perfect transmission, i.e., unity fidelity, for certain types of two-qubit errors.

Before closing this section, we mention a generalization of the above fidelity that will be needed in the presence of realistic single and two-qubit errors.  In general, single and two-qubit errors will lead to mixed input and output states with density matrices, $\tilde{\rho}_\text{I}^{\bm{r}}$ and $\tilde{\rho}_\text{O}$, respectively.  Here $\tilde{\rho}_\text{I}^{\bm{r}}=(\sigma^0+\bm{r}\cdot\hat{\sigma})/2$ and $\bm{r}$ is a Bloch sphere vector with $\vert \bm{r} \vert \leq 1$.  A mixed state fidelity is then given by \cite{Uhlmann1976,Jozsa1994,Liang2019}: $\left[\text{Tr}\left(\sqrt{ \sqrt{\tilde{\rho}_\text{I}^{\bm{r}}} U_{\Sigma} \tilde{\rho}_\text{O} U_{\Sigma} \sqrt{\tilde{\rho}_\text{I}^{\bm{r}}}  }\right)\right]^2$.  This generalization of fidelity reduces to the case we study in this paper, Eq.~(\ref{eq_fidelitysingle}), in the limit that the input state is pure, i.e., $\tilde{\rho}_\text{I}^{\bm{r}}\rightarrow \vert\psi_{\text{I}}^{\hat{r}}\rangle \langle \psi_\text{I}^{\hat{r}} \vert$.

The next sections describe how to build cluster states and measure fidelity.  Cluster states can be constructed from many different entangling gates.  But we use three different two-qubit gates built from three different types of interaction that we consider to be physical interactions: the controlled phase interaction, the Ising interaction, and the XY interaction.  We test fidelity measures on small cluster state chains to examine the impact of interaction errors on the fidelity.

\section{Cluster States from the Error-Prone Controlled Phase Interaction} \label{sec:controlled}

The controlled phase interaction between two qubits allows construction of cluster states with the fewest number of operations compared to the interactions considered in the following sections.  The controlled phase gate has also been realized with ionic \cite{Cirac1995,Ballance2016,Bruzewicz2019}, superconducting \cite{Strauch2003,Dicarlo2009}, and photonic \cite{Walther2005,Kok2007} qubits.  Furthermore, certain physical systems have native controlled phase interactions.  Rydberg interactions can directly implement the controlled phase interaction \cite{Jaksch2000,Lukin2001,Saffman2010,Saffman2005,Morgado2021}.  Also, controlled collisions \cite{Jaksch1999} using neutral atoms in hyperfine state-dependent optical lattices have realized parallel implementation of the controlled phase gate \cite{Mandel2003}.

Even though the controlled phase interaction requires the fewest gates to create a cluster state,  we will see that it has trade-offs in response to two-qubit error.  In this section, we will see that the controlled phase interaction does not allow preferred error-free teleportation channels.  In Sec.~\ref{sec:refocusing}, we will also see that it does not lend itself to simple refocusing using common schemes.

\begin{figure}
\begin{center}
\includegraphics[width=0.48\textwidth]{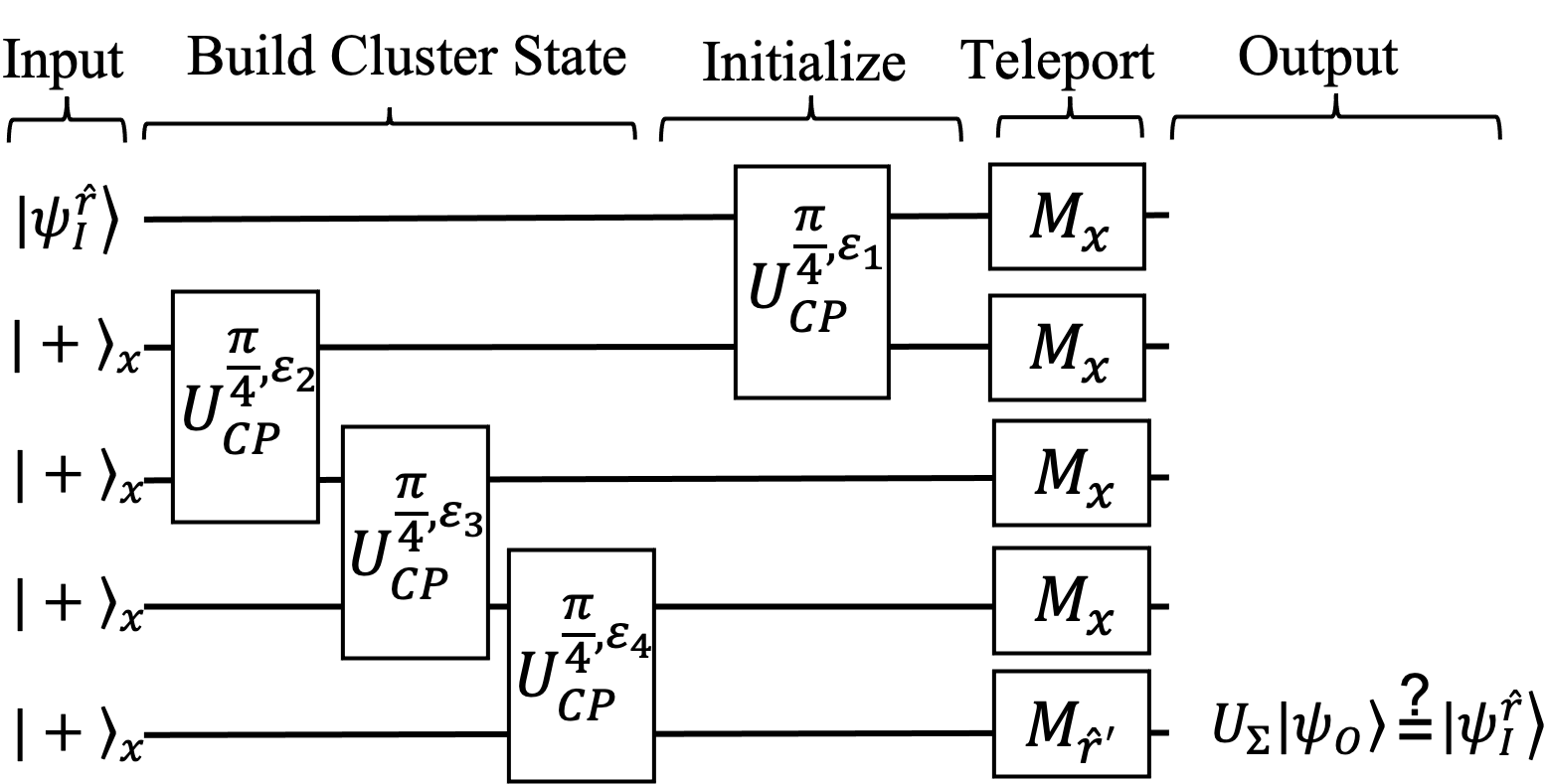}
\end{center}
\caption{
Circuit diagram depicting the construction of a 4-qubit cluster state.  The initial kets are entangled with the imperfect controlled phase interaction between qubits:  $U_{\text{CP}}^{\pi/4,\varepsilon}$, where $\varepsilon$ denotes inclusion of a dimensionless error parameter, Eq.~(\ref{eq_cphase_error}).  A fifth qubit, in state $\vert \psi_{\text{I}}^{\hat{r}} \rangle$, initializes information at one end of the cluster state.  The build and initialization protocols here consist of commuting operations and can therefore be performed in a different time sequence than the one shown, e.g., all at the same time. Single-qubit projective measurements along the qubit-$x$ direction, $\mathcal{P}(\text{M}_x)$, are recorded and used offline in $U_{\Sigma}$ [See Eqs.~(\ref{eq_Usigma_CP}) and (\ref{eq_fidelitysingle})].  The measurements teleport the information defined in $\vert \psi_{\text{I}}^{\hat{r}} \rangle$ along the chain to the last qubit where quantum state tomography yields a fidelity, Eq.~(\ref{eq_fidelitysingle}), which tends to unity as $\varepsilon\rightarrow0$.
}
\label{fig_cluster_gate_CP}
\end{figure}

The controlled phase interaction can be used to construct the two-qubit entangling interactions, the controlled phase shift gate.  We assume that the physical two-qubit interaction can be characterized by:
\begin{align}
H_{ij}^{\rm CP}=J^{\rm CP}_{ij} (\sigma^0_{i}-\sigma^z_{i})(\sigma^0_{j}-\sigma^z_{j}),
\end{align}
where $J^{\rm CP}_{ij}$ is the interaction energy between qubits $i$ and $j$ and is tunable in time.  $H_{ij}^{\rm CP}$ is useful for studying two-qubit error sources approximated by perturbations in $J^{\rm CP}_{ij}$.

To build the cluster state, we start with all qubits aligned along the $x$-direction:
\begin{align}
\vert \Phi_{0}^x \rangle =\prod_i \vert + \rangle_{x,i},
\end{align}
where $\vert \pm \rangle_{x,i}\equiv (\vert 0 \rangle_i \pm \vert 1\rangle_i)/\sqrt{2} $ are the eigenstates of $\sigma_i^x$.  Unitaries constructed from $H_{ij}^{\rm CP}$ can then be used to entangle qubits into the cluster state.  For the one-dimensional cluster state, we have \cite{Raussendorf2001b,Raussendorf2003}:
\begin{align}
\vert \Phi_\text{C} \rangle = \prod_{\langle i,j\rangle}  U_{\text{CP},i,j}^{\pi/4} \vert \Phi_{0}^x \rangle,
\label{eq_cluster_CP}
\end{align}
where $\langle i, j\rangle$ denotes a product over unique nearest neighbor pairs, i.e., $(1,2)$, $(2,3)$, ... .  Here we have used the controlled phase shift gate (the conditional phase gate) between qubits $i$ and $j$:
\begin{align}
U_{\text{CP},i,j}^{J_{i,j}^{\rm CP}t}\equiv e^{-i J_{i,j}^{\rm CP} t (\sigma^0_{i}-\sigma^z_{i})(\sigma^0_{j}-\sigma^z_{j}) },
\end{align}
where $\hbar$ is set to unity throughout. In matrix form, we see that the interaction appears in only one entry:
\begin{align}
U_{\text{CP},i,j}^{J_{i,j}^{\rm CP}t}=
\begin{pmatrix}
1 & 0 & 0 & 0\\
0 & 1 & 0 & 0\\
0 & 0 & 1 & 0 \\
0 & 0 & 0 & e^{-i4J^{\rm CP}_{i,j}t}
\end{pmatrix}
.
\label{eq_UCP_time}
\end{align}
Here and in the following, we construct matrices with the two-qubit basis of eigenstates of $\sigma_i^z\sigma_j^z$ in the following order: $\{ \vert +,+ \rangle_z,\vert +,- \rangle_z, \vert -,+ \rangle_z,\vert -,- \rangle_z \}$.

If we set $\theta=J^{\rm CP}_{i,j}t$ correctly, we can use the controlled phase interaction to realize the controlled-Z gate  (controlled phase-flip gate).  For $J^{\rm CP}_{i,j}t=\pi/4$, we have:
\begin{align}
U_{\text{CP},i,j}^{\pi/4}=
\begin{pmatrix}
1 & 0 & 0 & 0\\
0 & 1 & 0 & 0\\
0 & 0 & 1 & 0 \\
0 & 0 & 0 & -1 
\end{pmatrix}
.
\label{eq_CZ_matrix}
\end{align}

To study the impact of weak errors that are slow on the time scales of the two-qubit gates, we assume a static perturbation of the interaction strength.  We introduce the dimensionless perturbation $\varepsilon$ such that
\begin{align}
J^{\rm CP}_{i,j}t \rightarrow J^{\rm CP}_{i,j}t(1+\varepsilon).
\label{eq_cphase_error}
\end{align}
$\varepsilon$ parameterizes the unknown fractional deviation in the interaction strength due to slow error (or, equivalently, error in the pulse duration). This leads to the error-prone controlled phase shift gate (CPhase gate):
\begin{align}
U_{\text{CP}}^{\theta,\varepsilon}=
\begin{pmatrix}
1 & 0 & 0 & 0\\
0 & 1 & 0 & 0\\
0 & 0 & 1 & 0 \\
0 & 0 & 0 & e^{- i4J^{\rm CP}_{i,j}t (1+\varepsilon)} 
\end{pmatrix}
.
\label{eq_UCP_angle}
\end{align}

Figure~\ref{fig_cluster_gate_CP} shows the circuit diagram defining the construction of a small cluster state chain and subsequent measurement of all qubits needed to extract the fidelity.  As discussed in Sec.~\ref{sec_teleport}, to find the fidelity, we need to use $U_{\Sigma}$ to correctly interpret the basis for measurement of the output qubit.  After inserting  $U_{\Sigma}$, the results from quantum state tomography on the output qubit, and $\psi_{\text{I}}^{\hat{r}}$ into Eq.~(\ref{eq_fidelitysingle}), we can find the fidelity from measurements on an error-prone cluster state built from Eq.~(\ref{eq_UCP_angle}).  We now turn to numerical simulations to estimate the impact of error on the fidelity.

\begin{figure}[t]
\begin{center}
\includegraphics[width=0.46\textwidth]{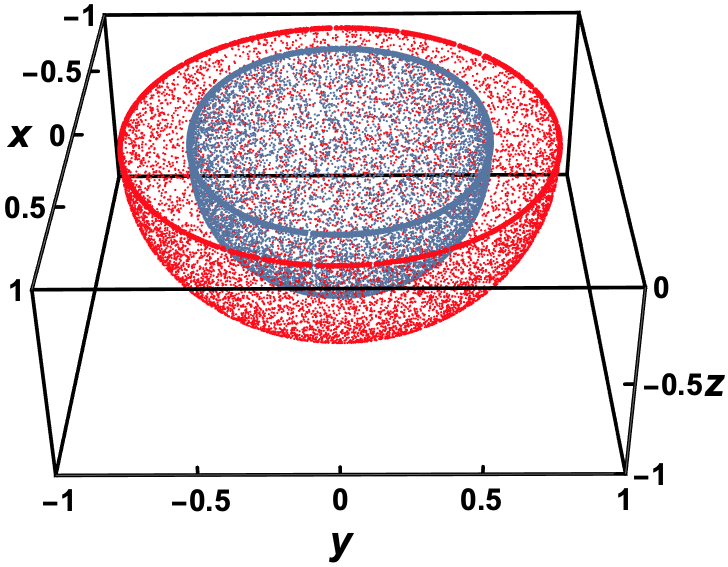}
\end{center}
\caption{
A cross section of the fidelity plotted for the  $N=5$ cluster state built from the error-prone controlled phase gate  $U_{\text{CP}}^{\pi/4,\varepsilon}$ as a function of error strength computed by assuming the same fixed error in all two-qubit gates ($\varepsilon_1=\varepsilon_2=\varepsilon_3=\varepsilon_4$ in Fig.~\ref{fig_cluster_gate_CP}).  The fidelity, Eq.~(\ref{eq_fidelitysingle}), is computed for the process described in Fig.~\ref{fig_cluster_gate_CP}.  We set a fixed interaction time using  $\theta=\pi/4$ in  Eq.~(\ref{eq_UCP_angle}),  so that the two-qubit interaction returns the desired controlled-Z gate (and therefore a perfect cluster state) in the absence of error.  The error is chosen as a static perturbation of the controlled phase interaction strength: $\pi/4(1+\varepsilon)$, so that $\varepsilon$ is the fractional change in the interaction strength. The initial state orientation $\hat{r}$ is chosen from a uniform distribution on the Bloch sphere to display benchmarking over varied initial states.  Here  the $x$, $y$, and $z$ axes correspond to  three respective directions in qubit space for the input qubit, $\hat{r}$.  The distance of the plotted points from the origin corresponds to the output fidelity, Eq.~(\ref{eq_fidelitysingle}).   For low error, e.g., $\varepsilon=1/(2\pi)$ (red circles), the output qubit is nearly identical to the input qubit leading to an approximate map of the Bloch sphere with nearly unity radius.  The blue circles correspond to a larger error, $\varepsilon=1/\pi$, where the fidelity is well below unity.  The resulting map of the Bloch sphere is a smaller shape. The resulting shape approximates a sphere but with weak anisotropy.}
\label{fig_3d_cphase}
\end{figure}

Errors will degrade the entanglement in the cluster state.  To study the role of error on the fidelity, we will assume that the single-qubit gates and all measurements are error-free, thus leaving error in just the controlled phase interaction.  We start by assuming that the two-qubit error is the same on all bonds. We will then, as a second step, randomize the two-qubit error as we move from qubit to qubit along the chain.

\begin{figure}[t]
\centering
\includegraphics[width=0.46\textwidth]{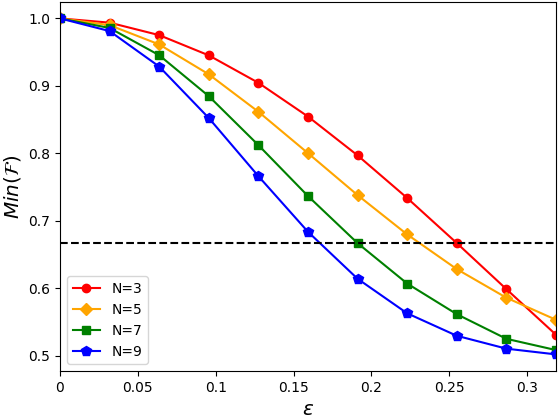}
\caption{The minimum fidelity of the cluster state built from the error-prone controlled phase gate  $U_{\text{CP}}^{\pi/4,\varepsilon}$ as a function of error strength computed by assuming the same fixed error in all two-qubit gates ($\varepsilon_1=\varepsilon_2=...=\varepsilon_N$ in Fig.~\ref{fig_cluster_gate_CP}). The fidelity, Eq.~(\ref{eq_fidelitysingle}), is computed by the process described in Fig.~\ref{fig_3d_cphase}. The minimum fidelity found from the initial state sampling distribution is chosen to show the worst case scenario.  Fidelities above the dashed line (2/3) teleport along a quantum channel.  As we increase the chain length from $N=3$ to $N=9$, the fidelity degrades. But we see that even the worst performing initial states allow teleportation through a quantum channel in cluster states defined with fractional errors in two-qubit gate strength as large as $15\%$.
}
\label{fig_minf_cp}
\end{figure}

To quantify the procedure depicted in Fig.~\ref{fig_cluster_gate_CP}, we first assume a fixed error $\varepsilon$ that is the same for all two-qubit gates.  We then vary the input qubit orientation, $\hat{r}$, using a uniform random distribution on the Bloch sphere for fixed $\varepsilon$.  We then find the orientation where the error makes the largest impact on the fidelity to compute $\text{Min}(\mathcal{F})$.

Figure~\ref{fig_3d_cphase} shows an example simulation of the fidelity for an $N=5$ cluster state for two different error strengths.  Each point plots a fidelity as measured from the origin.  The axes correspond to the orientation on the Bloch sphere of the input qubit, $\hat{r}$.  Here we see that for small error the fidelity nearly maps out a sphere of unit radius that corresponds to the Bloch sphere of the input qubit (red points). But larger error shrinks the sphere (blue points).  We therefore see that the error in Eq.~(\ref{eq_UCP_angle}) acts as a depolarizing channel insofar as the cluster state fidelity connects to the density matrix of the input qubit.

\begin{figure}[t]
\centering
\includegraphics[width=0.46\textwidth]{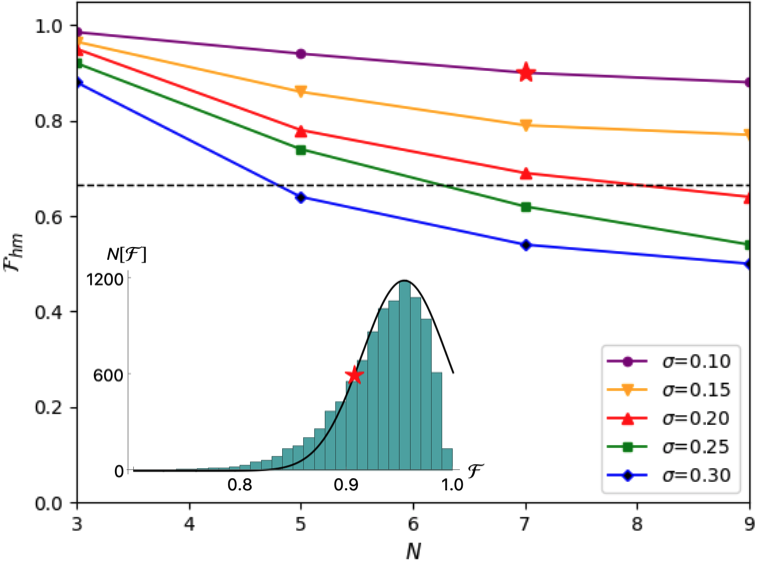}
\caption{ Fidelity averaged over randomized input qubit orientation and two-qubit disorder strengths that varies from bond to bond along the chain. Inset:  Histogram of fidelities obtained by constructing cluster states using the error-prone controlled phase interaction, $U_{\text{CP}}^{\pi/4,\varepsilon}$, with $\varepsilon$ a random variable chosen from a normal distribution of standard deviation $\sigma=0.1$ that disorders the two-qubit gate along the chain ($\varepsilon_1,\varepsilon_2,...,\varepsilon_N$ in Fig.~\ref{fig_cluster_gate_CP} are separate random variables).  The input qubit orientation is also randomly chosen, but from a uniform distribution of angles on the Bloch sphere.  The process depicted in Fig.~\ref{fig_cluster_gate_CP} is repeated for 7 qubits.  The histogram is truncated because the distribution of errors is still small enough to allow cases near the maximum fidelity ($\mathcal{F}=1$) for certain input states (See discussion in Sec.~\ref{sec:controlled}).  The star indicates the minimum half width at half maximum.  Main: Minimum half widths at half maximum of fidelities plotted as a function of the number of qubits for  various error distribution widths, $\sigma$.  The red star plots the same point as in the inset.  Fidelities above the dashed line are guaranteed to have used a quantum channel for teleportation.  We see that rather large errors are needed to pull the minimum fidelities below the dashed line for these small clusters states.
}
\label{fig_average_fidelity_cphase}
\end{figure}

We can derive closed formulae for the fidelity for small cluster states.  We take the initial state to have orientations $\hat{r}$ with the usual spherical coordinate angles $\theta_0$ and $\phi_0$ on the Bloch sphere, $\vert \psi_I^{\hat{r}=(\theta_0,\phi_0)}\rangle=\cos{(\theta_0/2)}\vert 0 \rangle+e^{i\phi_0}\sin{(\theta_0/2)} \vert 1 \rangle $.  For $N=3$, we find
\begin{align}
\mathcal{F}_{\hat{r}}\vert_{N=3}
& = 1 - (1 - \sin^2\theta_0\cos^2\phi_0)\sin^2{(\pi\varepsilon/2)}/2
\nonumber\\
& ~~~~ - \sin^2(\theta_0/2)\sin^2(\pi\varepsilon)/2.
\end{align}
This expression shows that the fidelity is only slightly anisotropic (nearly mapping out the uniform Bloch sphere). So, even though the minimum fidelity only appears for an initial state corresponding to a single point on the Bloch sphere, the minima can be approximately assumed to be at any angle.  The fidelity for large chains can be derived using cluster state refreshing for mathematical induction.  Appendix~\ref{sec:refresh} discusses refreshing that allows us to grow the cluster state by concatenation and therefore obtain analytic expressions for the fidelity for larger $N$.  We do not find a simple closed form for the minimum fidelity for larger $N$ for the controlled phase interaction, but will see that the Ising and XY interactions yield a simple closed form.  

Refreshing and concatenation of qubits (Appendix~\ref{sec:refresh}) can be used to systematically grow cluster states in experiments starting with the fewest number of qubits as possible.  One can use as few as three qubits with concatenation and refreshing to effectively grow the number of qubits used in teleportation.  However, in this limit, the measurement-based protocol is effectively the same as a circuit-based scheme to do the same (Appendix~\ref{sec:refresh}).  By systematically increasing the minimum number of qubits used to form the cluster state, e.g., as few as 5 qubits at once, one moves to the limit where teleportation along a cluster state chain begins to differ from the circuit-based scheme.

Figure~\ref{fig_minf_cp} plots the minimum fidelity as a function of the two-qubit error strength for several different cluster state chain lengths. The horizontal dashed line plots 2/3. The plot shows that the lowest fidelity is still above the threshold guaranteeing quantum teleportation (2/3) with perturbations to the interaction energy as large as $15\%$.  This assumes a cluster state with no more than 9 qubits.  This error is particularly large.  Our results for errors in the controlled phase interaction are consistent with the previous results \cite{Tame2005,TAME2006} that averaged over all input configurations for the controlled phase interaction.  The conclusion here is that one-dimensional cluster states allow a quantum channel for teleportation even for relatively large error strengths.  In the following sections, we will see that the Ising interaction allows simple analytic formulas to predict the scaling as we increase the chain length.

We now turn to numerical simulations which are more physically realistic in approximating how an experiment can map fidelity.  We assume random errors in all two-qubit gates used to build the cluster state such that  $\varepsilon$ is sampled from a normal distribution with standard deviation $\sigma$.  We repeat the calculation of fidelity for randomly selected input orientations, $\hat{r}$, and random $\varepsilon$ along the chain.  This type of simulation allows a complete benchmark using averaging over the input orientations as well as two-qubit error configurations. 

The inset in Fig.~\ref{fig_average_fidelity_cphase} shows an example histogram of fidelities for 7 qubits with $\sigma=0.1$. Here the Gaussian is truncated because the distribution hits the maximum fidelity.  For the controlled phase interaction, we find no states with perfect transmission in the presence of errors (unity fidelity). (The following section discusses interactions with perfect transmission in the cases of Ising and XY interactions.)  The unity fidelity here is due to the statistically significant likelihood of a low error configuration chosen from the normal distribution. The red star on the histogram denotes the half width minimum of all fidelities sampled. The lines in the main graph plot the half width minima found.  The star shown on the line is the same data point as highlighted by the star in the inset.  

Figure~\ref{fig_average_fidelity_cphase} shows that even for a broad distribution of two-qubit errors $\sigma \lesssim 0.15$, experiments can, on average, detect teleportation along a quantum channel.  This is consistent with robustness found for uniform error in Fig.~\ref{fig_minf_cp}.  But we also see that randomized error overall lowers the fidelity as one would expect.  The distribution tail in the inset of Fig.~\ref{fig_average_fidelity_cphase} even shows that some extreme cases of disorder drawn from our random sample of $\varepsilon$ strongly suppresses teleportation.

\section{Cluster States from the Error-Prone Ising Interaction}
\label{sec:Ising}

We now turn to cluster state chains built from the Ising interaction.  The Ising interaction characterizes NMR-based qubits \cite{Vandersypen2005} as well as superconducting charge-based qubits \cite{Krantz2019,You2005}.  The Ising gate also characterizes the M{\o}lmer-S{\o}renson gate between ions \cite{Molmer1999,Bruzewicz2019}. 

This section will show that the cluster state and measurement process for the Ising interaction are nearly identical to the same procedure as discussed for the controlled phase interaction in Sec.~\ref{sec:controlled}.  We therefore might expect that the cluster state fidelity responds in the same way.  We find similarities, but also find considerable differences in the fidelity and in refocusing schemes discussed later in Sec.~\ref{sec:refocusing}.

We first revisit the protocol introduced in Refs.~\cite{Raussendorf2001b,Raussendorf2003} to construct the cluster state from the Ising interaction.  We assume that the physical two-qubit interaction is given by:
\begin{align}
H_{ij}^{\rm ZZ}=J^{\rm Z}_{ij} \sigma^z_{i}\sigma^z_{j},
\label{eq_ising_H}
\end{align}
where $J^{\rm Z}_{ij}$ is the interaction energy between qubits $i$ and $j$ and is tunable in time.  To build the cluster state, we start with all qubits aligned along the $x$-direction $\vert \Phi_{0}^x \rangle$.  Unitaries constructed from $H_{ij}^{\rm ZZ}$ can then be used to build the cluster state:
\begin{align}
\vert \Phi_\text{C} \rangle = \prod_{\langle i,j\rangle} R_{\text{Z},i}^{-\pi/2 } R_{\text{Z},j}^{-\pi/2 } U_{\text{ZZ},i,j}^{\pi/4} \vert \Phi_{0}^x \rangle,
\label{eq_cluster_ZZ}
\end{align}
where
\begin{align}
U_{\text{ZZ},i,j}^{J_{i,j}^{\rm Z}t}\equiv e^{-i J_{i,j}^{\rm Z}t \sigma^z_{i}\sigma^z_{j} }.
\end{align}

The two qubit interaction can be recast in a matrix form. 
The matrix form for the Ising unitary is given by
\begin{align}
U_{\text{ZZ},i,j}^{J_{i,j}^{\rm Z}t}=
\begin{pmatrix}
e^{-iJ^{\rm Z}_{i,j}t } & 0 & 0 & 0\\
0 &e^{iJ^{\rm Z}_{i,j}t } & 0 & 0\\
0 & 0 & e^{iJ^{\rm Z}_{i,j}t } & 0 \\
0 & 0 & 0 & e^{-iJ^{\rm Z}_{i,j}t }
\end{pmatrix}
.
\end{align}
If we set $\theta=J^{\rm Z}_{i,j}t$ to parameterize time in units of the Ising interaction, then  $\theta=\pi/4$ defines the Ising gate needed to build the cluster state:
\begin{align}
U_{\text{ZZ},i,j}^{\pi/4}=e^{-i\pi/4}
\begin{pmatrix}
1 & 0 & 0 & 0\\
0 &i & 0 & 0\\
0 & 0 & i  & 0 \\
0 & 0 & 0 & 1 
\end{pmatrix}
.
\end{align}

We see from Eq.~(\ref{eq_cluster_ZZ}) that additional single qubit rotations are needed to build the cluster state that were not needed for the controlled phase interaction. 
A single qubit rotation about the $z$-axis is
\begin{align}
R_{\text{Z},i}^{\phi }=e^{-i\phi\sigma^z_{i}/2},
\end{align}
where $\phi$ parameterizes time evolution in units of the single qubit control field energy.

To study the impact of weak two-qubit errors that are slow on the time scales of all gates, we assume a static perturbation of the interaction strength,
\begin{align}
J^{\rm Z}_{i,j}t \rightarrow J^{\rm Z}_{i,j}t(1+\varepsilon).
\label{eq_error_Jz}
\end{align}
This type of error, as we will see below, leads to different behavior to the fidelity and will be a better estimate of the physically realistic sources of error in some qubit architectures.  These perturbations lead to the Ising gate with static error:
\begin{align}
U_{\text{ZZ}}^{\theta,\varepsilon}=e^{- i \theta (1+\varepsilon)\sigma^z_{i}\sigma^z_{j}}.
\label{eq_UZZ}
\end{align}
Unlike the controlled phase interaction, the error perturbs all non-zero entries in the Ising gate matrix.

\begin{figure}
\begin{center}
\includegraphics[width=0.48\textwidth]{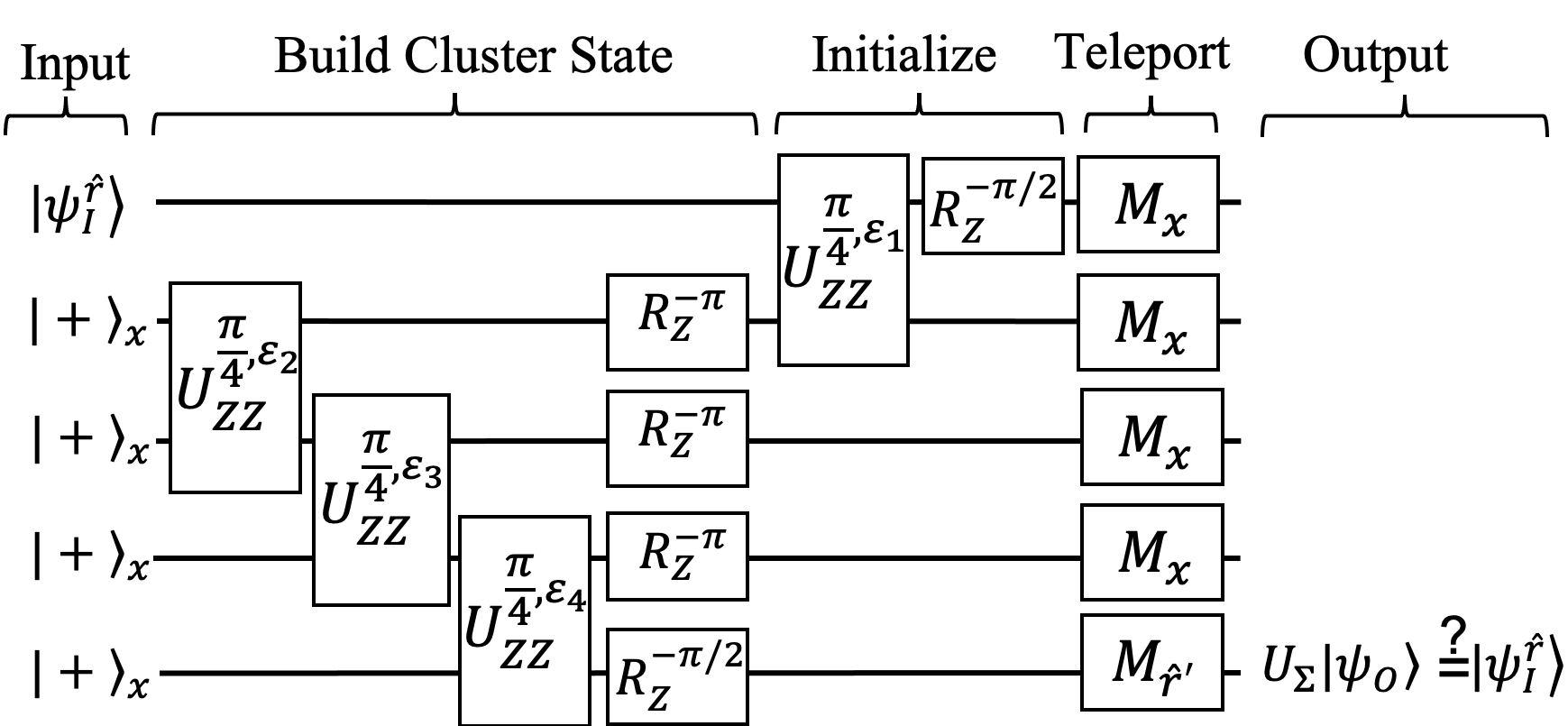}
\end{center}
\caption{Circuit diagram depicting the construction of a 4-qubit cluster state using the Ising interaction followed by an entanglement measure. The build and initialization protocols here consist of commuting operations and can therefore be performed in a different time sequence than the one shown here, e.g., all at the same time. The initial kets are entangled with the imperfect Ising interaction between two qubits,  $U_{\text{ZZ}}^{\pi/4,\varepsilon}$ where $\varepsilon$ is a dimensionless error parameter, Eq.~(\ref{eq_error_Jz}). $R_{\rm Z}^{\alpha}$ denotes single-qubit error-free rotation about the $z$-axis.  The series of measurements used to teleport the input state $\vert \psi_{\text{I}}^{\hat{r}} \rangle $ to the end of the chain with output $\vert \psi_{\text{O}} \rangle $ is the same as Fig.~\ref{fig_cluster_gate_CP}.  The single-qubit rotation gates can also be implemented by changing the measurement angles. }
\label{fig_cluster_gate_ZZ}
\end{figure}

Figure~\ref{fig_cluster_gate_ZZ} shows the circuit diagram for building the cluster state with the Ising gate and a subsequent measure of the fidelity. Figures~\ref{fig_cluster_gate_ZZ} and \ref{fig_cluster_gate_CP} differ in only two respects:  $i$) The single qubit rotations needed to convert the Ising gate to the controlled-Z gate and $ii$) the error-prone two-qubit interaction.  Otherwise the circuit diagrams show an identical procedure.  Nonetheless, the fidelity shows a different response to error.

\begin{figure}[t]
\begin{center}
\includegraphics[width=0.46\textwidth]{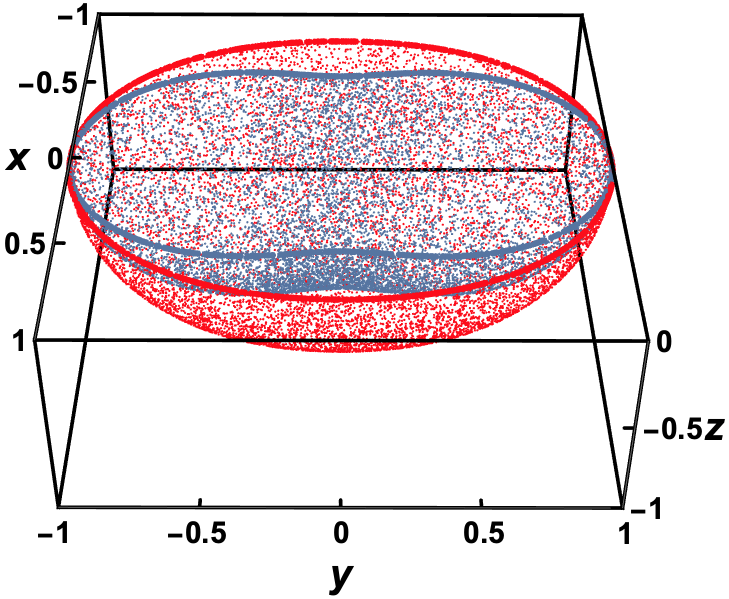}
\end{center}
\caption{
The same as Fig.~\ref{fig_3d_cphase} but for the Ising interaction. Red circles: $\varepsilon=1/\pi$, Blue circles: $\varepsilon=2/\pi$.  Here we also see that error in the two-qubit gate used to form the cluster state diminishes the fidelity.  But for initial qubit orientations $\hat{r}$ along the $\pm{\hat{y}}$ direction, we see that the error does \emph{not} impact the output result.  The corresponding fidelity for these initial states is unity and corresponds to perfect transmission along the cluster state in spite of two-qubit error. The resulting fidelity map is considerably distorted.  The minimum fidelity occurs for initial qubits oriented  anywhere in the $x$-$z$ plane, i.e., $\hat{r}\cdot\hat{y}=0$. The perfect transmission along $\pm{\hat{y}}$ can be used to diagnose the relative strength of Ising errors to all other errors. 
}
\label{fig_3d_Ising}
\end{figure}

As in Sec.~\ref{sec:controlled}, we randomly select the input state and assume uniform two-qubit error throughout the cluster state (the same error on all bonds). Figure~\ref{fig_3d_Ising} shows the same as  Fig.~\ref{fig_3d_cphase} but for the error-prone Ising interaction instead of the controlled phase interaction.  Remarkably, we see from the figure that the error does not impact the fidelity of states initially oriented along the $\pm \hat{y}$ direction on the Bloch sphere.  The maximum fidelity for the Ising interaction therefore corresponds to a case of perfect transmission along the chain in spite of Ising gate error. The maximum fidelity is: 
\begin{align}
\mathcal{F}_{\hat{r}=\pm\hat{y}}=1,
\end{align}
for arbitrary error.  We also find that the minimum fidelity occurs for initial qubits oriented in the $x$-$z$ plane, i.e., orientations such that $\hat{r}\cdot\hat{y}=0$. We therefore see that the error in Eq.~(\ref{eq_UZZ}) acts in a manner akin to a dephasing channel, in contrast to the depolarizing channel behavior seen for the controlled phase interaction.

The shape of Fig.~\ref{fig_3d_Ising} shows that the quality of quantum communication channels for the Ising interaction depends on the input state. This is due to the interplay of symmetry in the cluster state and the particular choice of Ising-gate error used in Eq.~(\ref{eq_UZZ}) (See Appendix~\ref{sec:perfect_transmission_perturbation} for a perturbative argument).   We note that perfect transmission with error-free single-qubit gates arises in a trivial unentangled limit, for $\varepsilon=-1$ because the single-qubit rotations alone are sufficient to leave $\mathcal{F}_{\hat{r}=\pm\hat{y}}=1$.  But it is surprising that, away from the trivial $\varepsilon=-1$ point, we have perfect transmission for $0<\varepsilon<1$, even where we have low (but non-zero) entanglement.  Future work will explore the role of symmetry in protecting these channels \cite{Zhang2021}.

\begin{figure}[t]
\centering
\includegraphics[width=0.46\textwidth]{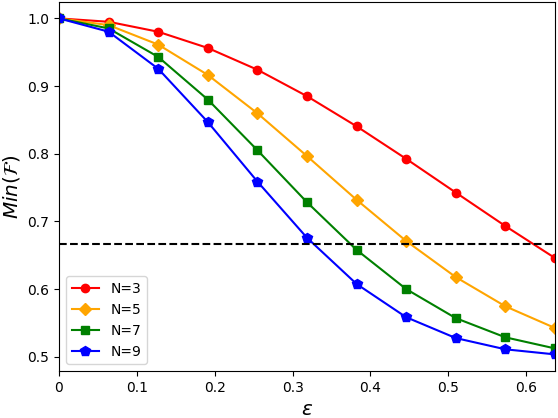}
\caption{
The same as Fig.~\ref{fig_minf_cp} but for the Ising interaction protocol described in Fig.~\ref{fig_cluster_gate_ZZ}.  Here we set the two-qubit gate time to be $J^{\text{Z}}t=\pi/4$ which leaves a perfect cluster state constructed from  Ising gates in the absence of error.  The minimum fidelity for the Ising interaction occurs for input qubits oriented in the $x$-$z$ plane, i.e.,  $\hat{r}\cdot\hat{y}=0$ in Fig.~\ref{fig_3d_Ising}. 
}
\label{fig_min_ising}
\end{figure}

Perfect transmission implies that, in experiments, we can use the fidelity to characterize departures from the error model captured by Eq.~(\ref{eq_error_Jz}).  If the error is only Ising error, then input qubits oriented along $\pm\hat{y}$ will reveal unity fidelity.  The presence of other error types, e.g., single-qubit error or measurement error, are therefore the only possible sources for departures from unity fidelity assuming $\hat{r}=\pm\hat{y}$.  Specifically, observing $\mathcal{F}_{\hat{r}=\pm\hat{y}}<1$ would measure all other types of error besides Ising errors.  $\mathcal{F}_{\hat{r}\cdot\hat{y}=0}$, by contrast, includes all types of errors.  By combining measurements to find the following ratio: $\mathcal{F}_{\hat{r}=\pm\hat{y}}/\mathcal{F}_{\hat{r}\cdot\hat{y}=0}$,
one measures the relative strength of the  non-Ising errors to all errors as they impact cluster state teleportation.  We therefore propose $\mathcal{F}_{\hat{r}=\pm\hat{y}}/\mathcal{F}_{\hat{r}\cdot\hat{y}=0}$ as a useful experimental diagnostic of error types.

The Ising interaction also allows the fidelity to be quantified by closed formulas.  We find that the minimum fidelity for the Ising interaction has a simple expression for uniform error on all gates defining the $N$-qubit cluster state chain.  As for the controlled phase interaction, we start with the $N=3$ chain fidelity for a qubit initially oriented along $\hat{r}=(\theta_0,\phi_0)$ on the Bloch sphere to find:
\begin{align}
\mathcal{F}_{\hat{r}}\vert_{N=3}
& = 1 - (1 -  \sin^2\theta_0\sin^2\phi_0)\sin^2(\pi\varepsilon/2)/2.
\label{fidelity_Ising_N3}
\end{align}
Here, we see that $\mathcal{F}_{\hat{r}=\pm\hat{y}}\vert_{N=3}=1$ for states along $\pm \hat{y}$, i.e., $\theta_0 = \pi/2$ and $\phi_0 = \pm \pi/2$.  But we also now see explicitly that there are minima for $\phi=0$ or $\pi$ for any $\theta$, i.e., the minima occurs in the $x$-$z$ plane with $\hat{r}\cdot y=0$.  We then find that $\text{Min}(\mathcal{F})\vert_{N=3}=\mathcal{F}_{\hat{r}\cdot\hat{y}=0}\vert_{N=3}=[3+\cos{(\pi\varepsilon)}]/4$.  The $N=3$ case can be extended to larger $N$ using refreshing and concatenation (see the Appendix~\ref{sec:refresh}).  For $N$ odd, we find
\begin{align}
\text{Min}(\mathcal{F}) = \frac{1 + \cos^{N-1}{(\pi\varepsilon/2)}}{2}.
\label{eq_ising_fidelity_analytic}
\end{align}
From this expression, we see that the maximum error allowing teleportation [obtained by setting $\text{Min}(\mathcal{F})=2/3$] is
\begin{align}
\varepsilon_{\text{max}}&=\frac{2}{\pi}\arccos{\left( 3^{\frac{1}{1-N}}\right)}
\nonumber\\
&=\sqrt{\frac{2\log3}{N}}+\mathcal{O}\big(N^{-3/2}\big).
\label{eq_emax}
\end{align}
This shows that significant two-qubit Ising errors can be tolerated in long cluster state chains since the fidelity scales as $1/\sqrt{N}$. We note, for comparison, that the wavefunction overlap between the error-free cluster state and the cluster state with Ising error is $\cos^{N-1}{(\pi\varepsilon/4)}$, which diminishes rapidly with increasing $N$.

\begin{figure}[t]
\centering
\includegraphics[width=0.46\textwidth]{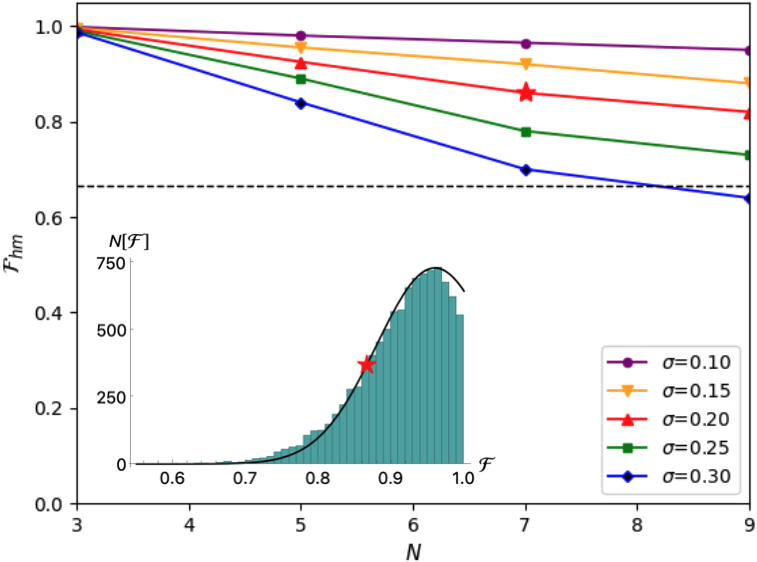}
\caption{The same as Fig.~\ref{fig_average_fidelity_cphase} but for the Ising interaction protocol described in Fig.~\ref{fig_cluster_gate_ZZ} and with the star chosen for $\sigma=0.2$.  Here the truncation of the Gaussian is due to the perfect transmission of two initial states through the disordered cluster state.  See Sec.~\ref{sec:Ising} for a discussion of perfect transmission.  
}
\label{fig_average_fidelity_ising}
\end{figure}

We now compare the fidelity in the presence of two-qubit Ising error to the results for the controlled-phase error by plotting them in the same manner.  We assume a benchmarking protocol identical to the one discussed in Sec.~\ref{sec:controlled}. To compare with the controlled phase interaction, Fig.~\ref{fig_min_ising} plots Eq.~(\ref{eq_ising_fidelity_analytic}).  By looking at where the solid lines cross the dashed line, we find the error $\varepsilon_{\text{max}}$ where the teleportation is no longer guaranteed to be along a quantum channel.  Here we also see that the minimum fidelity has the same qualitative behavior as shown for the controlled phase interaction in Fig.~\ref{fig_minf_cp}.

Figure~\ref{fig_average_fidelity_ising} plots the same as Fig.~\ref{fig_average_fidelity_cphase} but for the Ising interaction instead of the controlled phase interaction.   The inset shows a histogram that is truncated due to perfect transmission.  The truncation in the fidelity distribution is an observable qualitative difference between dominant error in the Ising interaction and the controlled phase interaction as we benchmark with the fidelity. As we sample the orientations of the input qubits with a uniform distribution, we find that the Gaussian distributed two-qubit errors still allow a significant number of cases of perfect transmission.

The main panel of Fig.~\ref{fig_average_fidelity_ising} appears to be qualitatively the same as Fig.~\ref{fig_average_fidelity_cphase} in spite of the cases of perfect transmission.  But quantitatively we see that the minimum fidelity is higher for the error-prone Ising interaction.  We therefore find that the minimum fidelity distribution reveals a somewhat more robust quantum channel for the error-prone Ising interaction than for the error-prone controlled phase  interaction.

\section{Cluster States from the Error-Prone XY Interaction}
\label{sec:XY}

\begin{figure}[t]
\begin{center}
\includegraphics[width=0.48\textwidth]{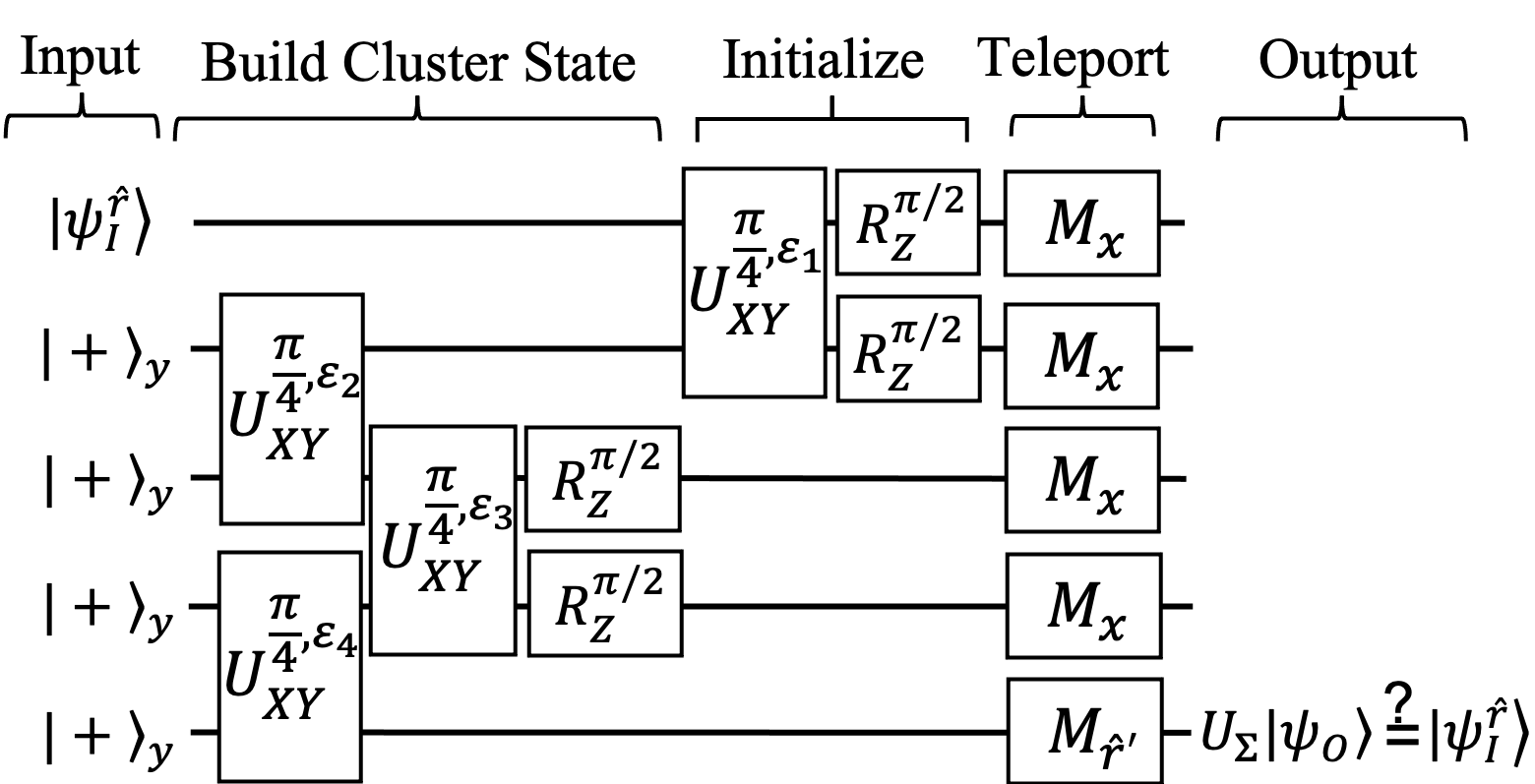}
\end{center}
\caption{Circuit diagram depicting the construction of a 4-qubit twisted cluster state using the XY interaction followed by an entanglement measure.  The twisted cluster state, Eq.~(\ref{eq_twisted_cluster_state}), is the same as the cluster state but with neighboring qubit labels on every other bond swapped. The initial kets denote four qubits prepared as eigenstates of $\sigma_y$, $\vert + \rangle_y$.  These four qubits are entangled with the imperfect XY interaction between two qubits,  $U_{\text{XY}}^{\pi/4,\varepsilon}$, where $\varepsilon$ is a dimensionless error parameter, Eq.~(\ref{eq_error_Jxy}).  The $U_{\text{XY}}^{\pi/4,\varepsilon}$ gates do not commute on bonds sharing a qubit and the order of execution must be respected, e.g., the XY interaction on all even bonds are executed simultaneously, followed by the simultaneous execution of the XY interaction on all odd bonds.  The series of measurements used to teleport the input state $\vert \psi_{\text{I}}^{\hat{r}} \rangle $ to the end of the chain with output $\vert \psi_{\text{O}} \rangle $ is the same as Fig.~\ref{fig_cluster_gate_CP}.  The single-qubit rotation gates can also be implemented by changing the measurement angles. 
}
\label{fig_cluster_gate_XY}
\end{figure}

We now turn to cluster states constructed from the error-prone XY interaction.  The XY interaction directly implements the iSWAP gate and characterizes interactions between qubits in several different qubit architectures.  Examples include the rotational states of a polar molecule placed in an optical tweezer trap  \cite{DeMille2002,Yan2013a,Ni2018}, quantum dots in cavities \cite{Imamo1999}, and certain types of superconducting qubits \cite{You2005}. 


We assume that the physical two-qubit interaction is given by
\begin{align}
H_{ij}^{\rm XY}=J^{\rm XY}_{ij} (\sigma^x_{i}\sigma^x_{j}+\sigma^y_{i}\sigma^y_{j}),
\label{eq_HXY}
\end{align}
where the interaction energy between qubits $i$ and $j$, $J^{\rm XY}_{ij}$, is tunable in time.  To build the cluster state we start with all qubits aligned along the $y$-direction:
\begin{align}
\vert \Phi_{0}^y \rangle =\prod_i \vert + \rangle_{y,i},
\end{align}
where $\vert \pm \rangle_{y,i}\equiv (\vert 0 \rangle_i \pm i \vert 1\rangle_i)/\sqrt{2} $ are the eigenstates of $\sigma_i^y$.

\begin{figure}[t]
\centering
\includegraphics[width=0.46\textwidth]{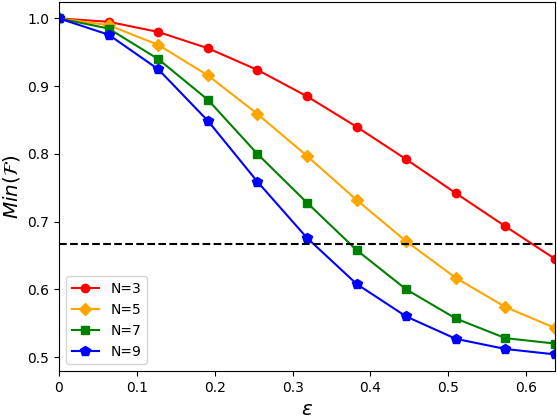}
\caption{
The same as Fig.~\ref{fig_minf_cp} but for the XY interaction protocol described in Fig.~\ref{fig_cluster_gate_XY}.  Here we set the two-qubit gate time to be $J^{\text{XY}}t=\pi/4$ which leaves a perfect cluster state constructed from iSWAP gates in the absence of error.
}
\label{fig_minf_XY}
\end{figure}

Unitaries constructed from $H_{ij}^{\rm XY}$ can be used to build the cluster state.  We have an efficient parallelizable protocol for building the ``twisted" cluster state \cite{Tanamoto2009a}:
\begin{align}
\vert \Phi_\text{C}^{\text{T}} \rangle = \prod_{i\neq 1} R_{\text{Z},i}^{\pi/2} \prod'_{i+1, i+2} U_{\text{XY},i+1,i+2}^{\pi/4} \prod'_{ i, i+1 } U_{\text{XY},i,i+1}^{\pi/4} \vert  \Phi_0^y \rangle,
\label{eq_twisted_cluster_state}
\end{align}
where the notation $\prod'_{ i, i+1}$ and $\prod'_{i+1, i+2}$ indicates a product over non-touching bonds $(i,i+1),(i+2,i+3),...$ and non-touching bonds $(i+1,i+2),(i+3,i+4),... $ , respectively.  $U_{\text{XY},ij}^{\theta} $ is given by:
\begin{align}
U_{\text{XY},ij}^{J^{\rm XY}_{ij}t}\equiv e^{-iJ^{\rm XY}_{ij}t  \left(\sigma^x_{i}\sigma^x_{j}+\sigma^y_{i}\sigma^y_{j}\right)}.
\end{align}
In matrix form, this becomes:
\begin{align}
U_{\text{XY},ij}^{J^{\rm XY}_{ij}t}=
\begin{pmatrix}
1 & 0 & 0 & 0\\
0 & \cos(2J^{\rm XY}_{ij}t) & -i\sin(2J^{\rm XY}_{ij}t) & 0\\
0 & -i\sin(2J^{\rm XY}_{ij}t) & \cos(2J^{\rm XY}_{ij}t) & 0 \\
0 & 0 & 0 & 1 
\end{pmatrix}
.
\end{align}
The twisted cluster state is the same as $\vert \Phi_C\rangle$ but with states at certain qubits swapped according to a twisting protocol \cite{Tanamoto2009a}.

To build the cluster state with the XY interaction, we relied on the iSWAP gate on qubits $i$ and $j$ with $J^{\rm XY}_{ij}t=\pi/4$:
\begin{align}
U_{\text{XY},ij}^{\pi/4}=
\begin{pmatrix}
1 & 0 & 0 & 0\\
0 & 0 & -i & 0\\
0 & -i & 0 & 0 \\
0 & 0 & 0 & 1 
\end{pmatrix}
.
\end{align}
But a faulty time evolution operator corresponding to the two-qubit interaction arises from the replacement
\begin{align}
J^{\text{XY}}_{i,j}t \rightarrow J^{\text{XY}}_{i,j}t(1+\varepsilon),
\label{eq_error_Jxy}
\end{align}
in Eq.~(\ref{eq_HXY}) and is given by:
\begin{align}
U_{\text{XY}}^{\theta,\varepsilon}=e^{- i \theta (1+\varepsilon)\left(\sigma^x_{1}\sigma^x_{2}+\sigma^y_{1}\sigma^y_{2}\right)},
\label{eq_UXY}
\end{align}
where we set $\theta=J^{\text{XY}}_{i,j} t$.  We will now examine the impact of these two-qubit XY errors on the cluster state fidelity.

\begin{figure}[t]
\centering
\includegraphics[width=0.46\textwidth]{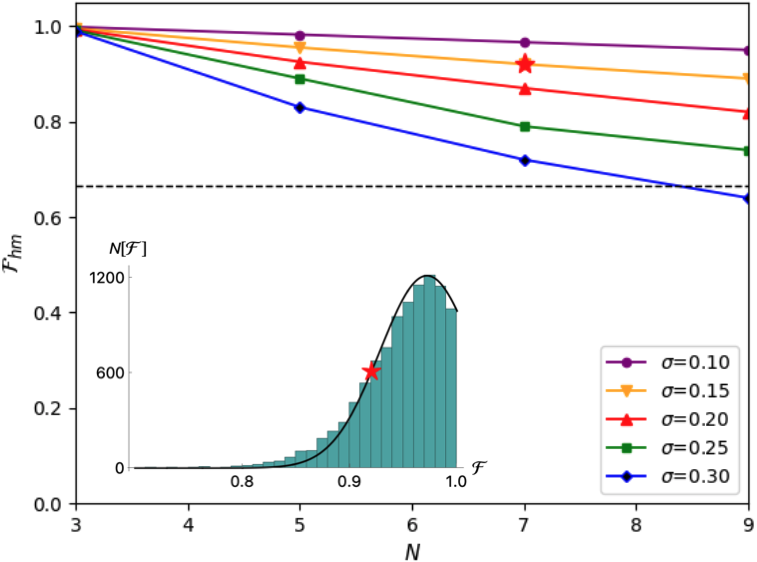}
\caption{
The same as Fig.~\ref{fig_average_fidelity_cphase} but for the XY interaction protocol described in Fig.~\ref{fig_cluster_gate_XY} and with the star chosen for $\sigma=0.15$.  Here, as in Fig.~\ref{fig_average_fidelity_ising},  the truncation of the Gaussian is due to the perfect transmission of two initial states through the disordered cluster state.
}
\label{fig_average_fidelity_XY}
\end{figure}

Figure~\ref{fig_cluster_gate_XY} depicts the circuit diagram needed to construct the cluster state with the error-prone XY interaction and measure the fidelity.   Note that the XY interaction on neighboring bonds does not commute, i.e., $[U_{\text{XY},ij}^{Jt},U_{\text{XY},jk}^{Jt}  ] \neq 0$, where $i\neq k$.  This results in two key differences between the protocol constructed for the Ising interaction and that for the XY interaction.  First, care must be taken in the pulse order to create the twisted cluster state as opposed to the Ising interaction.  Second, we note that the twisted cluster state can be thought of as the original cluster state with qubits $1,2,...,N-1$ swapped in neighboring pairs on every other bond.  We must therefore redefine the byproduct matrix $U_{\Sigma}$ . For the cluster state defined with the XY interaction, the $x$ and $z$ in Eq.~(\ref{eq_Usigma_CP}) are swapped.  With these two primary differences accounted for, the procedure for defining the cluster state with the XY interaction follows in the same manner as with the Ising interaction.

We simulated the process of measuring the fidelity of the cluster state formed from the error-prone XY interaction as in the previous sections.  We used the XY interaction and found results identical to Fig.~\ref{fig_3d_Ising} obtained using the error-prone Ising interaction.  For example, we find perfect transmission for $\hat{r}=\pm\hat{y}$ and the fidelity minima at $\hat{r}\cdot\hat{y}=0$.  We also find the same analytic expressions for the error bound on teleportation along quantum channels, Eqs.~(\ref{eq_ising_fidelity_analytic}) and ~(\ref{eq_emax}).

Figures~\ref{fig_minf_XY} and \ref{fig_average_fidelity_XY} quantify the response of the fidelity to errors in the XY interaction.  We find that the fidelity degrades in essentially the same manner as for the Ising interaction.  These results show that the protocol defined by Fig.~\ref{fig_cluster_gate_XY} puts the cluster state constructed from the XY interaction on the same footing as that constructed from the Ising interaction.

\section{Refocusing Pulses to Correct Errors in Two-qubit Gates}\label{sec:refocusing}

Cluster state sizes can be limited experimentally. Refreshing can be used to increase cluster state size in experiments with a limited number of qubits (See the Appendix~\ref{sec:refresh}).  We also saw in previous sections that, for fixed error strength, increasing the length of the cluster state eventually degrades the fidelity below 2/3.
Errors therefore limit the length of the cluster state chain that allows teleportation along a quantum channel.  If the dominant error sources are slow on time scales of two-qubit gates, we can use refocusing schemes to correct these gate errors \cite{Vandersypen2005}.  Refocusing is a powerful tool because it does not rely on a specific input state or on knowledge of the exact error strength.

By correcting slow gate errors, we can increase the length of the cluster state chains allowing teleportation along a quantum channel.  Refocusing schemes for single-qubit gate errors have been examined extensively \cite{Brown2004,Vandersypen2005}.  In what follows, we assume no error in single-qubit gates or in measurements.  We limit our analysis to refocusing schemes for slow two-qubit gate errors because these can dominate.  Such refocusing schemes can be used to correct errors to very high orders in $\varepsilon$  \cite{Brown2004,Hill2007} but at the expense of gate overhead.  We construct the simplest possible two-qubit refocusing pulse sequences to correct the errors studied above (Ising, XY, and controlled phase) to the lowest order.

\subsection{Refocusing Pulses for the Ising Interaction}
\label{sec_refocus_Ising}

We start with refocusing the error-prone Ising gate, Eq.~(\ref{eq_UZZ}).  Assuming an error-prone Ising interaction, $U_{\text{ZZ}}^{\theta,\varepsilon}$, we can construct a sequence of refocusing pulses to improve the accuracy of the two-qubit gate in approximating the exact Ising interaction, $U_{\text{ZZ}}^{\theta,0}$.  We construct sequences that simplify the two-qubit refocusing schemes following the usual BB1-type protocols of Refs.~\cite{Jones2003,Hill2007,Tomita2010a}. We use fewer pulses to refocus the interaction to only correct the lowest order errors (as opposed to the lowest and next-lowest order errors corrected by the longer BB1-type sequences \cite{Jones2003}).

The Ising gate, $U_{\text{ZZ}}^{\theta,\varepsilon}$, will have errors $\mathcal{O}(\varepsilon)$ that can be corrected to $\mathcal{O}(\varepsilon^2)$.  For any two qubits $i=1$ and $j=2$, we assume, in addition to the Ising interaction, Eq.~(\ref{eq_ising_H}), an error-free single-qubit control Hamiltonian: 
\begin{align}
H_{\text{X}}(B_x)_{1,2}=B_x\sigma^0_{1}\sigma^x_{2},
\label{eq_Hsingle_x}
\end{align}
which can be pulsed for a fixed duration in time leading to a propagator:
\begin{align}
U_{\text{X}}^{\delta}=e^{-i(\delta/2)\sigma_{1}^0\sigma_{2}^x},
\end{align}
where $\delta=2B_x t$.

The two Hamiltonians [Eqs.~(\ref{eq_ising_H}) and (\ref{eq_Hsingle_x})] lead to two different time scales $\theta$ and $\delta$ characterizing the interaction and single-qubit magnetic field, respectively.  They are related by
\begin{align}
\theta=\frac{ J^{\text{Z}}}{2B_x}\delta.
\end{align}
For convenience, we assume a time parameterization such that
\begin{align}
\frac{J^{\text{Z}}}{B_x}=\frac{8\pi\cos(\delta)}{\delta},
\end{align}
which can always be solved for at least one $\delta$.

\begin{figure}[t]
\begin{center}
\includegraphics[width=0.48\textwidth]{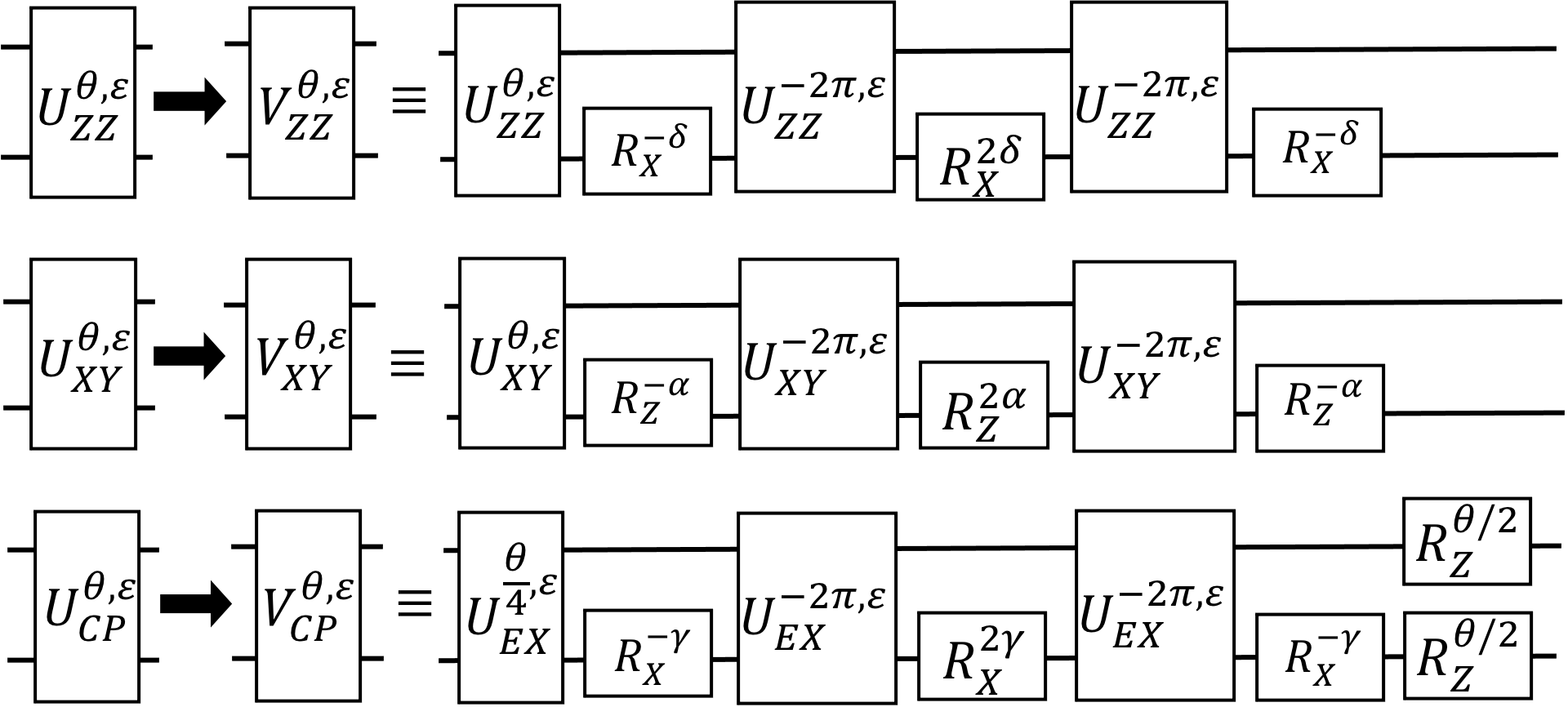}
\end{center}
\caption{Circuit diagrams depicting refocusing schemes used to improve faulty two-qubit gates under the assumption of perfect single-qubit gates. The top, middle, and bottom rows depict refocusing pulses for the error-prone Ising gate, Eq.~(\ref{eq_VZZ}), the XY gate, Eq.~(\ref{eq_VXY}), and the controlled phase gate, Eq.~(\ref{eq_VCP}), respectively. The controlled phase gate refocusing sequence is a composite sequence using $U_{\text{EX}}^{\theta,\varepsilon}$ with a further reduction depicted in Fig.~\ref{fig_refocusing_cphase}.
}
\label{fig_refocusing_ising_xy_cp}
\end{figure}

The central result of Ref.~\cite{Jones2003} was to show that known single-qubit refocusing schemes apply to errors in the two-qubit Ising gate as well.  Pulse sequences were shown to correct the \emph{two} lowest orders of errors. But we would like to construct a pulse sequence that is as short as possible and corrects only the lowest order. We find that the improved two-qubit Ising gate that corrects the lowest order of the error-prone Ising gate is
\begin{align}
V_{\text{ZZ}}^{\theta,\varepsilon}=
U_{\text{X}}^{-\delta}U_{\text{ZZ}}^{-2\pi,\varepsilon}(U_{\text{X}}^{-\delta})^{\dagger}
U_{\text{X}}^{\delta}U_{\text{ZZ}}^{-2\pi,\varepsilon}(U_{\text{X}}^{\delta})^{\dagger}
U_{\text{ZZ}}^{\theta,\varepsilon},
\label{eq_VZZ}
\end{align}
with an improved scaling in error:
$V_{\text{ZZ}}^{\theta,\varepsilon}=U_{\text{ZZ}}^{\theta,0}+\Delta U_{\text{ZZ}}\varepsilon^2+\mathcal{O}(\varepsilon^3)$.  Here the unwanted error is given by:
\begin{align}
\Delta U_{\text{ZZ}}=\begin{pmatrix}
0 & d_\delta & 0 & 0\\
-d_\delta^* & 0 & 0& 0\\
0 & 0& 0 & -d_\delta^*  \\
0 & 0 & d_\delta & 0 
\end{pmatrix},
\end{align}
with $d_{\delta}=-4\pi^2 i e^{4\pi i \cos{\delta}}\sin{(2\delta)}$.

The first row in Fig.~\ref{fig_refocusing_ising_xy_cp} depicts the circuit diagram used to create the refocused Ising gate, Eq.~(\ref{eq_VZZ}).  Replacing the two qubit gates in Fig.~\ref{fig_cluster_gate_ZZ} with $V_{\text{ZZ}}^{\theta,\varepsilon}$ from the first row in Fig.~\ref{fig_refocusing_ising_xy_cp} will improve the fidelity of the cluster state.  To characterize the improved two-qubit gate, we construct a two-qubit gate fidelity:
\begin{align}
F_2^{\rm ZZ}=\frac{\Big\vert \text{Tr}\left[V_{\text{ZZ}}^{\theta,\varepsilon}(U_{\text{ZZ}}^{\theta,0})^{\dagger}\right] \Big\vert}{\text{Tr}\left[U_{\text{ZZ}}^{\theta,0}(U_{\text{ZZ}}^{\theta,0})^{\dagger}\right]}.
\end{align}
The infidelity for the refocused gate, Eq.~(\ref{eq_VZZ}), is $1-F_2^{\rm ZZ}=8(\pi \varepsilon)^4\sin^2{(2\delta)}+\mathcal{O}(\varepsilon^6)$ which is improved by $\mathcal{O}(\varepsilon^2)$ over the infidelity for the original gate, Eq.~(\ref{eq_UZZ}),  $8(\pi\varepsilon)^2\cos^2{\delta}+\mathcal{O}(\varepsilon^4)$.  Longer pulse sequences can be constructed to further improve the fidelity \cite{Jones2003}.

\begin{figure*}[t]
\includegraphics[width=\textwidth]{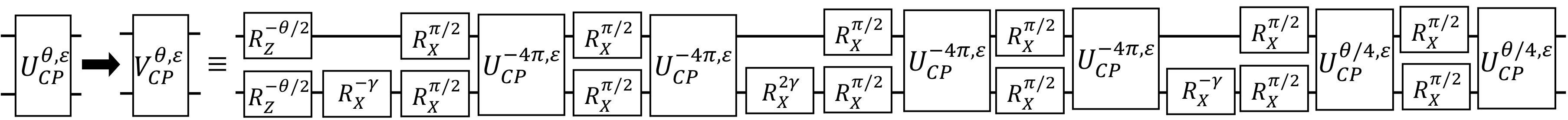}
\caption{Full circuit diagram depicting the refocusing scheme, Eqs.~(\ref{eq_UEX}) and (\ref{eq_VCP}), used to improve the faulty controlled phase gate.}
\label{fig_refocusing_cphase}
\end{figure*}

\subsection{Refocusing Pulses for the XY Interaction }
 
The errors in the XY gate, Eq.~(\ref{eq_UXY}), will lead to $\mathcal{O}(\varepsilon)$ errors as we attempt to construct the cluster state with the iSWAP gate.   These errors can also be corrected to $\mathcal{O}(\varepsilon^2)$, in a procedure similar to that for the Ising gate.  The goal of this section is to implement a good approximation to the exact XY time evolution operator,  $U_{\text{XY}}^{\theta,0}$, most importantly, the iSWAP gate,  $U_{\text{XY}}^{\pi/4,0}$, used in constructing the cluster state from the XY interaction.

We assume an error-free single-qubit control Hamiltonian:
\begin{align}
H_{\text{Z}}(B_z)_{1,2}=B_z\sigma^0_{1}\sigma^z_{2},
\label{eq_Hsingle_z}
\end{align}
where $H_{\text{Z}}$ is applicable for a fixed duration on qubits $i=1$ and $j=2$. The time evolution operator inducing single-qubit rotations of the second qubit about the Z-axis [obtained from pulsing Eq.~(\ref{eq_Hsingle_z})] is
\begin{align}
U_{\text{Z}}^{\alpha}=e^{-i(\alpha/2)\sigma^0_{1}\sigma^z_{2}},
\end{align}
where we use $\alpha$ to parameterize time
\begin{align}
B_zt=\alpha/2.
\label{eq_B}
\end{align}

The two Hamiltonians, Eqs.~(\ref{eq_HXY}) and (\ref{eq_Hsingle_z}), define two time scales.  The two time scales are related by
\begin{align}
\theta=\frac{ J^{\text{XY}}}{2B_z}\alpha.
\end{align}
For convenience, we assume a time parameterization such that
\begin{align}
\frac{J^{\text{XY}}}{B_z}=\frac{8\pi\cos(\alpha)}{\alpha},
\end{align}
which can always be solved for at least one $\alpha$.

To improve the accuracy of the two-qubit XY gate, we consider the refocused pulse sequence obtained in direct analogy to the Ising gate refocusing (Sec.~\ref{sec_refocus_Ising}).  It is given by
\begin{align}
V_{\text{XY}}^{\theta,\varepsilon}=
U_{\text{Z}}^{-\alpha}U_{\text{XY}}^{-2\pi,\varepsilon}(U_{\text{Z}}^{-\alpha})^{\dagger}
U_{\text{Z}}^{\alpha}U_{\text{XY}}^{-2\pi,\varepsilon}(U_{\text{Z}}^{\alpha})^{\dagger}
U_{\text{XY}}^{\theta,\varepsilon}.
\label{eq_VXY}
\end{align}
Here $V_{\text{XY}}^{\theta,\varepsilon}$ is an improved approximation to the iSWAP gate if we choose $\theta=\pi/4$.  The middle row in Fig.~\ref{fig_refocusing_ising_xy_cp} depicts the circuit diagram used to create the refocused XY gate.  Replacing the two qubit gates in Fig.~\ref{fig_cluster_gate_XY} with $V_{\text{XY}}^{\theta,\varepsilon}$ from the middle row in Fig.~\ref{fig_refocusing_ising_xy_cp} will improve the fidelity of the cluster state.

To see that $V_{\text{XY}}$ offers an improved approximation to the iSWAP gate, we expand in powers of $\varepsilon$:
$V_{\text{XY}}^{\theta,\varepsilon}=U_{\text{XY}}^{\theta,0}+\Delta U_{\text{XY}}\varepsilon^2+\mathcal{O}(\varepsilon^3)$ 
where the correction is given by
\begin{align}
\Delta U_{\text{XY}}=\begin{pmatrix}
0 & 0 & 0 & 0\\
0 & c_\alpha & -c'_\alpha & 0\\
0 & c'_\alpha & c^*_\alpha & 0 \\
0 & 0 & 0 & 0 
\end{pmatrix}
,
\end{align}
with $c_\alpha=i16 \pi^2 \cos[8\pi\cos(\alpha)] \sin(2\alpha)$
and $c'_\alpha=-16\pi^2$ $\times\sin[8\pi\cos(\alpha)]\sin(2\alpha)$. We then 
define the two-qubit gate fidelity:
\begin{align}
F_2^{\rm XY}=\frac{\Big\vert \text{Tr}\left[V_{\text{XY}}^{\theta,\varepsilon}(U_{\text{XY}}^{\theta,0})^{\dagger}\right] \Big\vert}{\text{Tr}\left[U_{\text{XY}}^{\theta,0}(U_{\text{XY}}^{\theta,0})^{\dagger}\right]}.
\end{align}
We find that the infidelity is then $1-F_2^{\rm XY}=64(\pi\varepsilon)^4$ $\times\sin^2(2\alpha)+\mathcal{O}(\varepsilon^6)$.  This shows that refocusing to implement $V_{\text{XY}}$ offers an $\mathcal{O}(\varepsilon^2)$ improvement to the faulty iSWAP gate since the infidelity of the un-refocused pulse, Eq.~(\ref{eq_UXY}), is much larger, specifically, $16(\pi\varepsilon)^2\cos^2(\alpha)$.

\subsection{Refocusing Pulses for the  Controlled Phase Interaction}

Refocusing of the controlled phase interaction can, in principle, use the method constructed for the Ising interaction above, because the Ising and controlled phase gates are equivalent up to single-qubit rotations.  But these added single-qubit rotations imply the need for an additional extraction procedure \cite{Hill2007} which adds to the gate overhead.  

To construct the refocusing scheme for the controlled phase interaction, we assume that the physical two-qubit interaction is given by $H_{ij}^{\rm CP}$  acting on qubits $i$ and $j$ leading to the propagator $U_{\text{CP}}^{\theta,\varepsilon}$
where $\theta=4J^{\rm CP}_{i,j}t$ for any time $t$.  But we must now also assume two different error-free single-qubit control Hamiltonians.  First we assume that
\begin{align}
H_{\text{X}}(B_x)_{i,j}=B_x\sigma^0_{i}\sigma^x_{j}
\label{eq_Hsingle_x_cp}
\end{align}
can be pulsed for a fixed duration in time leading to a propagator $U_{\text{X}}^{\gamma}$ where $\gamma=2B_x t$. For convenience, we set $\gamma=\arccos{(\theta/16\pi)}$.  We also assume the single-qubit control Hamiltonian is given by
\begin{align}
H_{\text{Z}}(B_z)_{i,j}=B_z\sigma^0_{i}\sigma^z_{j},
\label{eq_Hsingle_z_cp}
\end{align}
leading to a propagator $U_{\text{Z}}^{\gamma'}$
where $\gamma'=2B_z t$ .

To refocus and improve the accuracy of the controlled phase gate, we rely on the pulse sequence for the Ising interaction. But we must extract \cite{Jones2003,Hill2007} the Ising term from the controlled phase interaction by noting 
$U_{\text{CP}}^{\theta,0}=e^{-i\theta}e^{i\theta(\sigma^0_{i}\sigma^z_{j}+\sigma^z_{i}\sigma^0_{j})}U_{\text{ZZ}}^{\theta,0} $.  The Ising term can be isolated using
$e^{i\frac{\pi}{2}(\sigma^0_{i}\sigma^x_{j}+\sigma^x_{i}\sigma^0_{j})}
$ since these rotations essentially remove the role of the leading $Z$ rotations:
\begin{align}
U_{\text{EX}}^{\theta,\varepsilon}\equiv
e^{-i\frac{\pi}{2}(\sigma^0_{i}\sigma^x_{j}+\sigma^x_{i}\sigma^0_{j})}
U_{\text{CP}}^{2\theta,\varepsilon}
e^{-i\frac{\pi}{2}(\sigma^0_{i}\sigma^x_{j}+\sigma^x_{i}\sigma^0_{j})}
U_{\text{CP}}^{2\theta,\varepsilon},
\label{eq_UEX}
\end{align} 
where the Ising term is extracted from the controlled phase interaction
(up to a phase of $e^{i\theta(1+\varepsilon)}$).

We can now use $U_{\text{EX}}^{\theta,\varepsilon}$ in place of the Ising pulse in the refocusing sequence discussed in Sec.~\ref{sec_refocus_Ising}.  The composite refocusing sequence for the controlled phase interaction then becomes nearly the same as Eq.~(\ref{eq_VZZ}):
\begin{align}
V_{\text{CP}}^{\theta,\varepsilon}
&= 
e^{i\frac{\theta}{4}(\sigma^0_{i}\sigma^z_{j}+\sigma^z_{i}\sigma^0_{j})}
U_{\text{X}}^{-\gamma}U_{\text{EX}}^{-2\pi,\varepsilon}
\nonumber \\
&~~~\times (U_{\text{X}}^{-\gamma})^{\dagger}
U_{\text{X}}^{\gamma}U_{\text{EX}}^{-2\pi,\varepsilon}(U _{\text{X}}^{\gamma})^{\dagger}
U_{\text{EX}}^{\theta/4,\varepsilon},
\label{eq_VCP}
\end{align}
where the primary difference is the leading exponential term which reinserts the prefactors needed for the corrected controlled phase gate.

The bottom row in Fig.~\ref{fig_refocusing_ising_xy_cp} shows the circuit diagram for $V_{\text{CP}}^{\theta,\varepsilon}$ written in terms of the composite pulse Eq.~(\ref{eq_UEX}) for comparison with the Ising and XY sequences.  Note that extracted gates, Eq.~(\ref{eq_UEX}), are themselves composite gates.  Figure~\ref{fig_refocusing_cphase} depicts Eq.~(\ref{eq_VCP}) in a full circuit diagram written in terms of the physical controlled phase gate rather than Eq.~(\ref{eq_UEX}).

Replacing $U_{\text{CP}}^{\theta,\varepsilon}$ with $V_{\text{CP}}^{\theta,\varepsilon}$ in Fig.~\ref{fig_cluster_gate_CP} will reduce error in constructing the controlled phase gate.  To quantify the improvement, we define the two-qubit gate fidelity:
\begin{align}
F_2^{\rm CP} = \frac{\Big\vert \text{Tr}\Big[V_{\text{CP}}^{\theta,\varepsilon}(U_{\text{CP}}^{\theta,0})^{\dagger}\Big] \Big\vert}{\text{Tr}\Big[U_{\text{CP}}^{\theta,0}(U_{\text{CP}}^{\theta,0})^{\dagger}\Big]}.
\end{align}
The corrected gate sequence then has an infidelity $1-F_2^{\rm CP}=- \frac{\theta^2(\theta^2-256\pi^2)}{2048} \varepsilon^4 + \mathcal{O}(\varepsilon^6)$ which improves over the infidelity of $U_{\text{CP}}^{\theta,\varepsilon}$, which is $3\varepsilon^2\theta^2/32 + \mathcal{O}(\varepsilon^4)$.  The improved two-qubit fidelity for the controlled phase interactions shows that we can replace the two qubit gates in Fig.~\ref{fig_cluster_gate_CP} with the corrected sequence from Fig.~\ref{fig_refocusing_cphase}. While this should, in principle, help improve the overall cluster state fidelity, we see that the overall gate count for the refocused XY and Ising interaction are much smaller than that for the controlled phase interaction.  Further compactifications of these gate sequences could shorten them.

\section{Summary}\label{sec:summary}

We have systematically analyzed the construction of cluster state chains from faulty interactions along with a cluster state teleportation-based fidelity measure of entanglement.  We focused on the controlled phase, Ising, and XY interactions.  We find that errors in different interaction strengths lead to different fidelity responses.   By running numerical experiments designed to simulate key pieces of experimental benchmarking, we find that errors in the Ising and XY interaction have a preferred direction in qubit space.  We find, in particular, a case of perfect transmission in spite of errors in the Ising and XY interaction. The fidelity discussed here can therefore be used as a measure of interaction error in comparison to all other errors.   We also find that the $1/\sqrt{N}$ scaling of maximum errors for teleportation along a quantum channel leaves room for growing cluster state chains while preserving entanglement.

We have also discussed refocusing schemes for improving the two-qubit gate fidelities in the presence of slow two-qubit gate errors.  The refocusing schemes discussed here can be made more compact, extended to correct error to higher orders \cite{Jones2003,Hill2007}, and can also be paired with single-qubit refocusing \cite{Tomita2010a}.  Our work sets the stage for combining randomized benchmarking with refocusing to experimentally grow and test entanglement on cluster states.

\begin{acknowledgments}
We thank R. Raussendorf and E. Sela for helpful discussions. All authors acknowledge support from the Air Force Office of Scientific Research Grant No. FA9550-19-1-0272. S.K. acknowledges funding from the Army Research Office Grant No. W911NF-17-1-0563. Z.Q., V.S., and W.L. acknowledge support by the Air Force Office of Scientific Research Grant No. FA9550-18-1-0505 and Army Research Office Grant No W911NF-20-1-0013.
\end{acknowledgments}

\bibliography{references}

\appendix

\section{Qubit Refreshing}
\label{sec:refresh}

\begin{figure}[t]
\begin{center}
\includegraphics[width=0.38\textwidth]{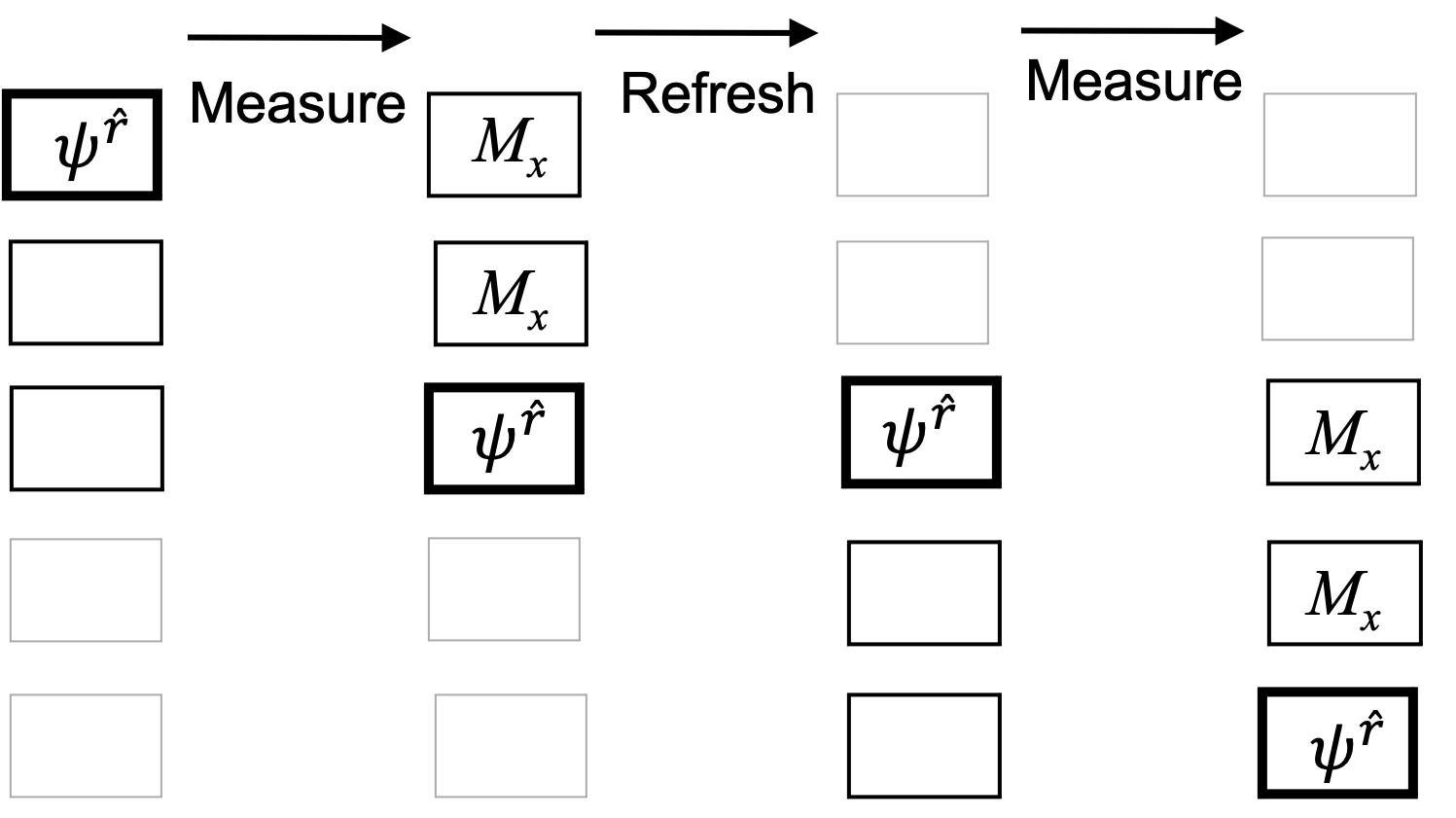}
\end{center}
\caption{Schematic depicting qubit refreshing of a 5-qubit cluster state as we teleport $\vert \psi_{\hat{r}} \rangle $ using measurements on at most 3 qubits.  The bold boxes contain the information regarding $\vert \psi_{\hat{r}} \rangle$.  The greyed-out qubits are not needed. The first column depicts the initial qubit prepared in the state $\vert \psi_{\hat{r}} \rangle $ and entangled with two other qubits in a cluster state.  The second column depicts measurements that move $\vert \psi_{\hat{r}} \rangle $ from the first to the third qubit.  The third column depicts the entangling of two new qubits into the cluster state after the first two qubits are discarded.  The final row depicts the final measurements that moves $\vert \psi_{\hat{r}} \rangle $ to the last qubit.  The entire refreshing process depicted here uses at most 3 qubits at the same time while effectively teleporting along a 5-qubit cluster state.
}
\label{fig_refreshing}
\end{figure}

Qubit refreshing for cluster states describes the process by which qubits are recycled after the measurement process \cite{Raussendorf2001b,Raussendorf2003}.  Given a cluster state, measurements on a set of qubits collapse only the part of the cluster state wavefunction corresponding to the measured qubits. This implies that those qubits can be removed and re-entangled elsewhere on the graph.  Fig.~\ref{fig_refreshing} shows a schematic of the refreshing process for teleportation (MBQC identity gate) along a cluster state chain.  The example in the figure shows that only 3 qubits are needed at any one time to effectively teleport along a 5-qubit cluster state chain.  More generally, refreshing shows that only 3 qubits are needed for measuring fidelity along a chain of arbitrary length.

We relied on refreshing in the main text. We used it in our derivations where, in  Sec.~\ref{sec:controlled}, we used refreshing to derive  Eq.~(\ref{eq_ising_fidelity_analytic}).  We also relied on refreshing to argue that it can also be used experimentally.  If experiments are limited in qubit resources, refreshing can be used to build larger cluster states.  As the minimum number of qubits used increases, the process of MBQC teleportation along the chain begins to differ from the circuit-based scheme.  In MBQC, more single qubit measurements are done after the application of the two-qubit gates.  This is to be compared with the circuit-based scheme where two-qubit gates are applied throughout the algorithm rather than upfront.

\section{Perfect transmission for weak error}
\label{sec:perfect_transmission_perturbation}

In this section, we show analytically that, for $\varepsilon \ll 1$, the cluster state exhibits perfect transmission for input qubit aligned along the $y$ direction for the Ising interaction (similarly for the XY interaction).  We also show that this is not the case for the controlled phase interaction. For this purpose, it is sufficient to consider just the $N=3$ cluster state.  Using refreshing (Appendix~\ref{sec:refresh}), one can extend the argument here to larger $N$. 

We first focus on the error-prone Ising interaction.  We make a perturbation expansion in $\varepsilon$:
\begin{align}
U_{{\rm ZZ},i,j}^{\pi/4,\varepsilon} 
& = U_{{\rm ZZ},i,j}^{\pi/4,0} 
\sum_{n=0}^\infty \Gamma_{{\rm ZZ},i,j}^{(n)} \varepsilon^n,
\end{align}
where the coefficient matrices are defined by
\begin{align}
\Gamma_{{\rm ZZ},i,j}^{(n)}
& = \frac{1}{n!} \left(-i\frac{\pi}{4}\right)^n \left\{
\begin{array}{cl}
\sigma_i^0\sigma_j^0, & n\in\textrm{even},
\\
\sigma_i^z \sigma_j^z, & n\in\textrm{odd}.
\end{array}
\right.
\end{align}
Accordingly, the cluster state is expanded as:
\begin{align}
|\psi_I^{\hat{r}}, \Phi_C\rangle_\varepsilon
& = \prod_{\langle i,j\rangle} \left(\sum_{n=0}^\infty \Gamma_{{\rm ZZ},i,j}^{(n)} \varepsilon^n \right) |\psi_I^{\hat{r}}, \Phi_C\rangle_0.
\end{align}
For the density matrix $\rho_\varepsilon(\psi_I^{\hat{r}}, \Phi_C) = |\psi_I^{\hat{r}}, \Phi_C\rangle_\varepsilon\langle\psi_I^{\hat{r}}, \Phi_C|$, the perturbation expansion reads:
\begin{align}
\rho_\varepsilon(\psi_I^{\hat{r}}, \Phi_C) 
& = \rho_0(\psi_I^{\hat{r}}, \Phi_C) + \sum_{n=1}^\infty \tilde{\rho}_{\rm ZZ}^{(n)}(\psi_I^{\hat{r}}, \Phi_C) \varepsilon^n.
\end{align}
Here, we are interested in the first two coefficient matrices ($n=1,2$) given by
\begin{align}
\tilde{\rho}_{\rm ZZ}^{(1)}(\psi_I^{\hat{r}}, \Phi_C)
& = \tilde{\Gamma}_{\rm ZZ}^{(1)} \rho_0(\psi_I^{\hat{r}}, \Phi_C) + \rho_0(\psi_I^{\hat{r}}, \Phi_C) \tilde{\Gamma}_{\rm ZZ}^{(1)\dag},
\\
\tilde{\rho}_{\rm ZZ}^{(2)}(\psi_I^{\hat{r}}, \Phi_C)
& = \tilde{\Gamma}_{\rm ZZ}^{(2)} \rho_0(\psi_I^{\hat{r}}, \Phi_C) + \rho_0(\psi_I^{\hat{r}}, \Phi_C) \tilde{\Gamma}_{\rm ZZ}^{(2)\dag}
\nonumber\\
& ~~~~ + \tilde{\Gamma}_{\rm ZZ}^{(1)} \rho_0(\psi_I^{\hat{r}}, \Phi_C) \tilde{\Gamma}_{\rm ZZ}^{(1)\dag},
\label{density_matrix_2nd_order_Ising}
\end{align}
where, for $N=3$, we define
\begin{align}
\tilde{\Gamma}_{\rm ZZ}^{(1)}
& = \Gamma_{{\rm ZZ},1,2}^{(1)} \Gamma_{{\rm ZZ},2,3}^{(0)} + \Gamma_{{\rm ZZ},1,2}^{(0)} \Gamma_{{\rm ZZ},2,3}^{(1)}
\nonumber\\
& = -i\frac{\pi}{4} (\sigma_1^z \sigma_3^0 + \sigma_1^0 \sigma_3^z) \sigma_2^z,
\\
\tilde{\Gamma}_{\rm ZZ}^{(2)}
& = \Gamma_{{\rm ZZ},1,2}^{(2)} \Gamma_{{\rm ZZ},2,3}^{(0)} + \Gamma_{{\rm ZZ},1,2}^{(0)} \Gamma_{{\rm ZZ},2,3}^{(2)} + \Gamma_{{\rm ZZ},1,2}^{(1)} \Gamma_{{\rm ZZ},2,3}^{(1)}
\nonumber\\
& = -\frac{\pi^2}{16} (\sigma_1^0 \sigma_3^0 + \sigma_1^z \sigma_3^z) \sigma_2^0.
\end{align}
It can be shown that, for a qubit initially oriented along $\hat{r}=(\theta_0,\phi_0)$ on the Bloch sphere, $\vert \psi_I^{\hat{r}=(\theta_0,\phi_0)}\rangle=\cos{(\theta_0/2)}\vert 0 \rangle+e^{i\phi_0}\sin{(\theta_0/2)} \vert 1 \rangle$, the output density matrix and the fidelity have the following closed form (up to second order in $\varepsilon$):
\begin{align}
\rho_O|_{N=3}
& = \rho_I^{\hat{r}} + 
\tilde{\rho}_O^{(2)} \varepsilon^2 + \mathcal{O}(\varepsilon^3),
\\
\mathcal{F}_{\hat{r}}|_{N=3}
& = 1 + \tilde{\mathcal{F}}_{\hat{r}}^{(2)} \varepsilon^2 + \mathcal{O}(\varepsilon^3),
\end{align}
where the leading correction arises in the second order:
\begin{align}
\tilde{\rho}_O^{(2)} 
& = -\frac{\pi^2}{8}\big[\rho_I^{\hat{r}=(\theta_0,\phi_0)} - \rho_I^{\hat{r}'=(\pi-\theta_0,\pi-\phi_0)}\big],
\label{Output_2nd_order_Ising}
\\
\tilde{\mathcal{F}}_{\hat{r}}^{(2)}
& = - \frac{\pi^2}{8}\left(1 - \sin^2\theta_0 \sin^2\phi_0\right).
\label{Fidelity_2nd_order_Ising}
\end{align}

We can use Eqs.~\eqref{Output_2nd_order_Ising} and \eqref{Fidelity_2nd_order_Ising} to show perfect transmission. The first term of Eq.~\eqref{Output_2nd_order_Ising} does not perturb the direction of the input qubit, giving an angle-independent contribution to Eq.~\eqref{Fidelity_2nd_order_Ising}. The second term of Eq.~\eqref{Output_2nd_order_Ising} describes a qubit oriented along $\hat{r}' = (\pi-\theta_0,\pi-\phi_0)$ (preserving the sum $\hat{r}+\hat{r}'$), giving rise to the angle dependence in Eq.~\eqref{Fidelity_2nd_order_Ising}. For the symmetric choice of $\hat{r} = \hat{r}' =  (\pi/2,\pm\pi/2)$, two terms in Eqs.~\eqref{Output_2nd_order_Ising} and \eqref{Fidelity_2nd_order_Ising} cancel each other, leading to perfect transmission in the second order. The same argument is expected to be applied to the higher orders as already shown in Eq.~\eqref{fidelity_Ising_N3}.

For comparison, the perturbation expansion for the controlled phase interaction is also made as follows:
\begin{align}
U_{{\rm CP},i,j}^{\pi/4,\varepsilon} 
& = U_{{\rm CP},i,j}^{\pi/4,0} 
\sum_{n=0}^\infty \Gamma_{{\rm CP},i,j}^{(n)} \varepsilon^n,
\end{align}
where the coefficient matrices are defined by
\begin{align}
\Gamma_{{\rm CP},i,j}^{(n)}
& = \frac{(-i\pi)^n}{4n!} (\sigma_i^0 - \sigma_i^z) (\sigma_j^0 - \sigma_j^z).
\end{align}
As before, the leading correction to the fidelity starts from the second order:
\begin{align}
\tilde{\mathcal{F}}_{\hat{r}}^{(2)}
= - \frac{\pi^2}{8}\left[1 - \sin^2\theta_0 \cos^2\phi_0 + 4\sin^2(\theta_0/2)\right].
\label{Fidelity_2nd_order_CP}
\end{align}
Here, we see that Eq.~\eqref{Fidelity_2nd_order_CP} always gives a negative contribution for any $\theta_0$ and $\phi_0$ in contrast to Eq.~\eqref{Fidelity_2nd_order_Ising}.  The controlled phase interaction therefore does not allow perfect transmission.

\end{document}